\documentclass[fleqn,usenatbib]{mnras}

\usepackage{newtxtext,newtxmath}

\usepackage[T1]{fontenc}
\usepackage{ae,aecompl}

\usepackage{graphicx}	
\usepackage{amsmath}	

\usepackage{enumitem}
\setlist[itemize]{leftmargin=*}

\usepackage{epsfig}
\usepackage{multirow}
\usepackage{float}
\usepackage{footmisc}

\def\apjl{ApJL}
\def\apjs{ApJS}
\def\aap{A\&A}

\def\xmm{{\it XMM-Newton}}
\def\cxo{{\it Chandra}}

\title[Optical/UV and X-ray Correlations of Quasars]{Wavelength Dependences of the Optical/UV and X-ray Luminosity Correlations of Quasars}

\author[C. Jin, et al.]{Chichuan Jin$^{1,2}$\thanks{E-mail: ccjin@nao.cas.cn},
Elisabeta Lusso$^{3,4}$,
Martin Ward$^{5}$,
Chris Done$^{5}$,
Riccardo Middei$^{6,7}$
\smallskip
\\
$^{1}$National Astronomical Observatories, Chinese Academy of Sciences, 20A Datun Road, Beijing 100101, China\\
$^{2}$School of Astronomy and Space Sciences, University of Chinese Academy of Sciences, 19A Yuquan Road, Beijing 100049, China\\
$^{3}$Dipartimento di Fisica e Astronomia, Università di Firenze, Via G.\ Sansone 1, 50019 Sesto Fiorentino, FI, Italy\\
$^{4}$INAF - Osservatorio Astrofisico di Arcetri, Largo E.\ Fermi 5, 50125 Firenze, Italy\\
$^{5}$Centre for Extragalactic Astronomy, Department of Physics, University of Durham, South Road, Durham DH1 3LE, UK\\
$^{6}$Space Science Data Center, SSDC, ASI, via del Politecnico snc, 00133 Roma, Italy\\
$^{7}$INAF-Osservatorio Astronomico di Roma, via Frascati 33, 00078 Monteporzio Catone, Italy
}

\date{prepared for MNRAS}

\pubyear{2022}

\begin{document}
\label{firstpage}
\pagerange{\pageref{firstpage}--\pageref{lastpage}}
\maketitle

\begin{abstract}
The inter-band correlations between optical/UV and X-ray luminosities of active galactic nuclei (AGN) are important for understanding the disc-coronal connection, as well as using AGN as standard candles for cosmology. It is conventional to measure the X-ray luminosity at rest frame 2~keV and compare to the UV luminosity at the rest-frame 2500 \AA, but the wavelength-dependence was never well explored. In this work, we adopt a well-defined sample of 1169 unobscured quasars in the redshift range 0.13 -- 4.51, and apply the direct-correlation method to explore how the correlation with the 2 keV luminosity changes at different optical/UV wavelengths, from 1280 -- 5550 \AA\ where the spectral quality is high. We find that the luminosity at all UV continuum wavelengths correlates with the X-ray luminosity similarly to that at 2500 \AA, and that these correlations are better than at the optical wavelengths. Strong self-correlation is also found in the broadband optical/UV continuum, supporting the scenario that it is dominated by the disc emission. Correlations of various emission lines are also investigated (e.g. C {\sc iv},  C {\sc iii}], Mg {\sc ii}, H$\beta$, [O {\sc iii}]$\lambda\lambda 4959/5007$), including the Baldwin effect and correlations involving line-widths. We find the forms of these line correlations are different, and they are also different from their underlying continua, suggesting various complexities in the line-generation process. We discuss these results in the disc-wind scenario. Our study confirms that the rest-frame 2500 \AA\ is a good wavelength to represent the optical/UV continual properties of quasars, and shows the advantages of the direct-correlation method.


\end{abstract}




\section{Introduction}
\label{sec-intro}
The relationship observed between the X-rays and optical/UV emission of active galactic nuclei (AGN) has been studied for decades (e.g. \citealt{Tananbaum.1979, Vignali.2003, Just.2007, Lusso.2010,Jin.2012c,Vagnetti.2013, Lusso.2016, Risaliti.2019, Lusso.2020, Bisogni.2021}). Observationally, these two quantities are correlated with a slope of $\simeq0.6$ and an observed dispersion that varies from $0.4-0.2$ dex depending on the sample selection.
From a physical perspective, it is generally accepted that the primary hard X-ray emission arises from a compact hot corona (e.g. \citealt{Mushotzky.1993, Zdziarski.1996, Zdziarski.1999}) around the super-massive black hole, while the optical/UV continuum mainly comes from the outer region of the accretion disc (\citealt{Shakura.1973, Novikov.1973}). 
However, no adequate physical model able to explain such a relation between the X-ray and optical/UV emission exists yet (i.e. the physical connection between the disc and warm/hot corona in unknown; see the recent review by \citealt{Liu.2022}).

This non-linear UV/X-ray luminosity correlation is also important because it can be used to study cosmology \citep[e.g.][]{Risaliti.2015,Risaliti.2019}. Quasars can be employed as standard candles to measure distances (in particular at $z > 2$) and to constrain cosmological parameters (e.g. \citealt{melia2019,bargiacchi2022}). Furthermore, it has been verified, across a wide redshift range, that the slope of the UV/X-ray luminosity correlation do not show any statistically significant redshift evolution up to redshift of about 6 (see e.g. \citealt{Salvestrini.2019,Lusso.2020}) and that the dispersion of the UV/X-ray luminosity correlation is mainly caused by measurement uncertainties and some intrinsic variability (e.g. \citealt{Lusso.2016, Chiaraluce.2018}).

The adoption of the luminosity at 2500 \AA\ is somewhat arbitrary. This choice can be traced back to Maarten Schmidt's seminal paper in 1968 (\citealt{Schmidt.1968}). Their motivations for employing 2500 \AA\ were that this wavelength is not affected by emission lines and lies in the middle of wavelength range (i.e. 1700 -- 3500 \AA) where the continuum has a power law shape. Additionally, the optical/UV continuum depends on the mass, mass accretion rate and spin of the black hole (\citealt{Shakura.1973, Novikov.1973}), and is also affected by other components such as dust extinction (e.g. \citealt{Fitzpatrick.1999, Fitzpatrick.2007}) and UV iron lines (e.g. \citealt{Verner.1999, Vestergaard.2001}). A fundamental question is thus whether 2500 \AA\ can effectively be a good proxy for the properties of optical/UV continuum of quasars in general, especially whether its correlation with X-rays is representative. Therefore, to provide a better understanding of the physics driving the X-ray-to-UV relation, one should analyse whether the luminosity correlation between the optical/UV and X-rays has a wavelength dependence \citep[e.g.][]{young2010}. Recently, \citet[][]{signorini23} performed a complete UV spectroscopic analysis of a sample of $\simeq$1800 quasars with SDSS optical spectra and \xmm\ X-ray serendipitous observations. In the X-rays, they analysed the spectra of all the sample objects at redshift $z > 1.9$, while they considered catalogued photometric measurements at lower redshifts. They find that the monochromatic fluxes at 1 keV and 2500 \AA\ are, respectively, the best X-ray and UV continuum indicators among those that are typically available. However, the best-fit slope that they obtained, by employing spectroscopic UV fluxes and analysing the X-ray-to-UV relation in narrow redshift bins, is somewhat flatter ($\simeq0.46$) than what observed with photometric values ($\simeq0.57$, see their Figure 3).

In addition, by correlating the optical/UV data with the X-ray luminosity, we can study the origin of the optical/UV continuum and various emission/absorption lines. \citet{Jin.2012b} correlated the 2--10 keV luminosity with the optical luminosity for a sample of 51 AGN with Sloan Digital Sky Survey (SDSS) spectra, and derived the variation of the correlation coefficient with wavelength, namely the optical-to-X-ray correlation spectrum (OXCS). The main advantage of OXCS is that it is not model-dependent (i.e. does not depend on the modelling of the continuum and line profile), so it can directly reflect the intrinsic correlations of the continuum and emission lines (including individual line components) with X-rays. The OXCS reported by \citet{Jin.2012b} revealed various correlation features. For example, the strongest correlation was found between hard X-rays and [O {\sc iii}], which is stronger than the continuum and emission lines at other optical wavelengths. The H$\beta$ broad line component correlates more strongly with X-rays than with the narrow line component. These results provide important clues for our understanding of the relationship between different emission line regions and their ionization sources. However, the sample in \citet{Jin.2012b} is relatively small (51 AGN) and is limited to $z < 0.4$, so their SDSS spectra cannot cover UV in the rest-frame. Therefore, it is necessary to use larger quasar samples covering wider ranges of redshifts. In addition, we can replace the X-ray luminosity with UV luminosity, in order to examine the self-correlation of the emission in the optical/UV band. This UV self-correlation can be used to distinguish different optical/UV continual components, and also to better understand the origins of optical/UV lines.

Large quasar samples at various redshifts with well-calibrated SDSS spectra offer a good opportunity to address the questions mentioned above. Hence, the main motivation of this work is to investigate the wavelength dependence of the optical/UV and X-ray luminosity correlations using a large sample of SDSS quasars, and so to test the basis for the choice of 2500 \AA. Also, we extend the OXCS of \citet{Jin.2012b} into the UV band, and explore the UV self-correlation spectrum. These results improve our understanding of the origin of the optical/UV continuum and various emission lines.

The structure of this paper is as follows. First, we describe the sample and its statistical properties in Section~\ref{sec-obs}. Then in Section~\ref{sec-optuvx} we show the optical/UV and X-ray correlation spectra and the optical/UV self-correlation spectra. Section~\ref{sec-wdep} presents the wavelength dependence of the regression parameters of the optical/UV and X-ray luminosity correlations. Discussion about the choice of 2500 \AA, further correlation analyses and the origin of various optical/UV emission lines are presented in Section~\ref{sec-discussion}. The main results of this work are summarized in the final section. A flat universe model with H$_{0} = 72$ km s$^{-1}$ Mpc$^{-1}$, $\Omega_{\Lambda} = 0.73$, $\Omega_{\rm M} = 0.27$ is adopted throughout the paper.

\begin{figure*}
\centering
\begin{tabular}{cc}
\includegraphics[trim=0.5in 0.1in 0.0in 0.0in, clip=1, scale=0.55]{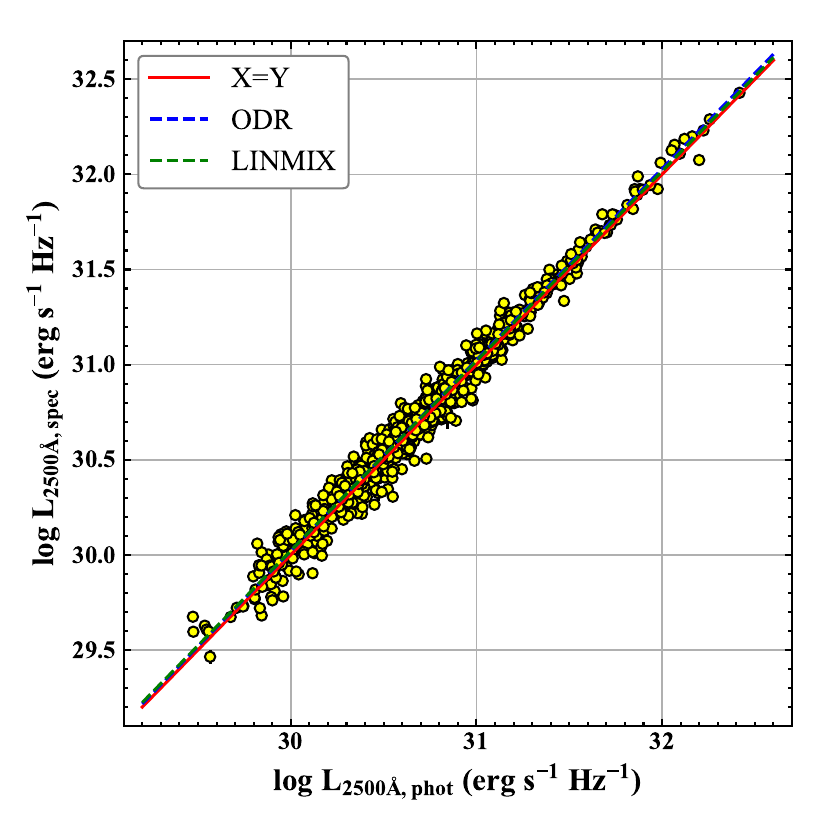}  &
\includegraphics[trim=-0.5in 0.1in 0.0in 0.0in, clip=1, scale=0.55]{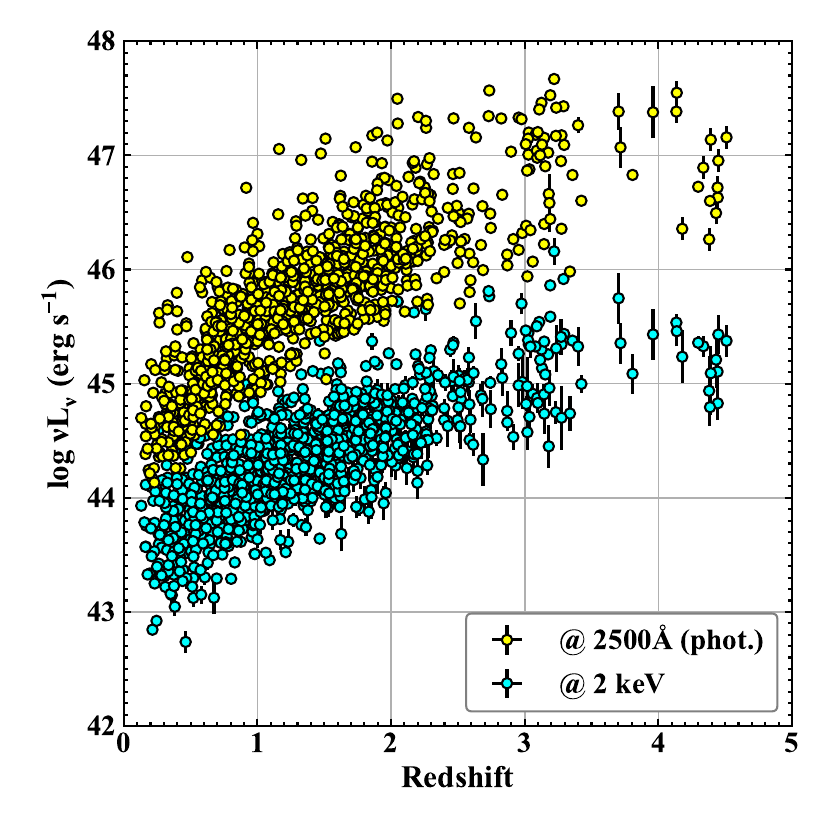}  \\
\end{tabular}
\caption{Left panel: cross-correlating the photometric luminosity at 2500 \AA\ with the spectral luminosity at 2500 \AA\ for a subsample of the 902 quasars whose SDSS spectra cover the rest-frame 2500 \AA. The correlation is fully consistent with the $X=Y$ relation. The intrinsic scatter is 6.6 per cent. Right panel: redshift distribution of the monochromatic luminosities at 2500 \AA\ (photometry, yellow) and 2 keV (cyan) for the entire sample of 1169 quasars.}
\label{fig-2500cross}
\end{figure*}

\section{The Quasar Sample}
\label{sec-obs}
\subsection{Sample Selection}
The study of the optical/UV and X-ray luminosity correlation requires a large, unobscured quasar sample with high-quality optical/UV and X-ray data. The parent sample of this work is taken from \citet{Lusso.2020}, which contains 2421 quasars with optical spectroscopy from the Sloan Digital Sky Survey Data Release 14 (SDSS DR14) and X-ray observations from either \xmm\ or \cxo. This parent sample is one of the cleanest quasar samples, in the sense that all quasars with strong extinction (i.e. dust reddening and X-ray absorption), host galaxy contamination and Eddington bias have been excluded (see \citealt{Lusso.2020} for more details). Since the aim of this work is to explore the wavelength dependence of the inter-band correlation, it is necessary to use the SDSS spectra with the best absolute flux calibration. 
In fact, there is a known issue with the flux calibration of quasar spectroscopy in DR12 and prior data releases \citep[see e.g.][for details]{milakovic2021}. Briefly, BOSS quasar targets are observed with an offset on the focal plane to increase the S/N around the Ly$\alpha$, but since the standard stars are not observed in the same fashion, the derived flux correction is not suited for quasars. This results in an overall bluer quasar spectra. A correction has been developed and applied to the individual exposures and the final coadded spectra in DR14 and later data releases, which improved the flux calibration, although there are still significant residuals\footnote{\url{https://www.sdss3.org/dr9/spectro/caveats.php}} \citep[e.g.][see their Section~2]{ws2022}.
We thus crossmatch the parent sample with the SDSS DR7 catalogue, finding 1278 unique sources. All of these sources have the {\tt sciencePrimary} flag equal to 1, indicating that the spectra are good for scientific analysis\footnote{\url{https://skyserver.sdss.org/dr7/en/help/docs/glossary.asp}}. There are 128 sources having multiple SDSS spectra, for which we choose the spectrum with the highest signal-to-noise (S/N) for our subsequent analysis. 
We then visually inspect all the SDSS spectra and further exclude 105 sources whose spectra contain data gaps, or if the averaged S/N is $<$ 10 (i.e. \textsc{sn\_median\_all}>10, which is the S/N per observation over the entire wavelength range and for all observations), or the continuum shows a suspicious down-turn towards the blue end of the spectrum which is most likely due  to reddening. The final clean sample contains 1169 quasars. 

\subsection{Data Preparation}
The SDSS DR7 spectra were downloaded from the SDSS Data Archive Server\footnote{\url{http://das.sdss.org/www/html/das2.html}}. For the purpose of investigating the detailed wavelength dependence of various correlation parameters, we must rebin all de-redshifted SDSS spectra on common wavelengths. A relatively low spectral resolution of $\lambda/\Delta\lambda=200$ is adopted to increase the S/N in each wavelength bin. Using a larger $\lambda/\Delta\lambda$ will mainly result in more noisy spectra, but will not change the results of this work. The wavelength coverage of SDSS is 3800 -- 9200 \AA, but the instrument throughput decreases towards the edges\footnote{\url{http://classic.sdss.org/dr7/instruments/spectrographs/index.html}} where the S/N also decreases, and so we restrict the spectral range to 4000 -- 8600 \AA\ for the following analysis. All spectra are de-reddened for Galactic extinction along their line of sight, using the dust map of \citet{Schlegel.1998} and the reddening curve of \citet{Fitzpatrick.2007}, and then de-redshifted to their respective individual rest-frames.

The catalogue table of \citet{Lusso.2020}\footnote{\url{https://cdsarc.cds.unistra.fr/viz-bin/cat/J/A+A/642/A150}} contains the X-ray flux at 2 keV (rest-frame) and the UV flux at 2500 \AA\ from photometry. However, the photometric flux may include contamination from both emission and absorption lines (e.g. strong UV Fe {\sc ii} lines) in the filter's bandpass, so we considered the spectral flux to study the correlation in our analysis. We then check whether there is statistical agreement between the photometric and the spectral flux values.
We first convert all the fluxes to luminosities, and then cross-correlate the photometric luminosity with the spectral luminosity at 2500 \AA. Note that only 902 sources in our sample are used for this analysis, because these all lie in the redshift range of 0.60 -- 2.44 (corresponding to 4000--8600 \AA), and so the SDSS spectra can the rest-frame 2500 \AA.

As shown in Figure~\ref{fig-2500cross} left panel, the two estimates of luminosities at 2500 \AA\ are very consistent with each other. We apply two regression algorithms to check the correlation. The first is the orthogonal distance regression (ODR) algorithm, which treats the two variables symmetrically. The second is the LINMIX algorithm\footnote{\url{https://linmix.readthedocs.io/en/latest/src/linmix.html}} (\citealt{Kelly.2007}), which uses Monte Carlo Markov Chains (MCMC) and returns a reliable estimate of the intrinsic scatter. Both algorithms take into account variable uncertainties (see \citealt{Lusso.2016} for more details). The ODR algorithm obtains a slope of 1.0046 $\pm$ 0.0044 and an intercept of -0.119 $\pm$ 0.136 (blue dash line), and the LINMIX algorithm finds a slope of 0.998 $\pm$ 0.005 and an intercept of 0.094 $\pm$ 0.148 (green dash line). Both are consistent with the $X=Y$ relation (red solid line). The intrinsic scatter is found to be 6.63 $\pm$ 0.17 per cent. These results confirm that the photometric and spectral fluxes are all acceptably well calibrated, and the contamination by nearby emission (or absorption) lines in the photometric flux is not significant. Therefore our SDSS spectra can be used to study the wavelength dependence of the correlations, and the results can then be compared with previous studies based on the photometric flux at 2500 \AA.

\begin{figure}
\centering
\includegraphics[trim=0.0in 0.2in 0.0in 0.1in, clip=1, scale=0.55]{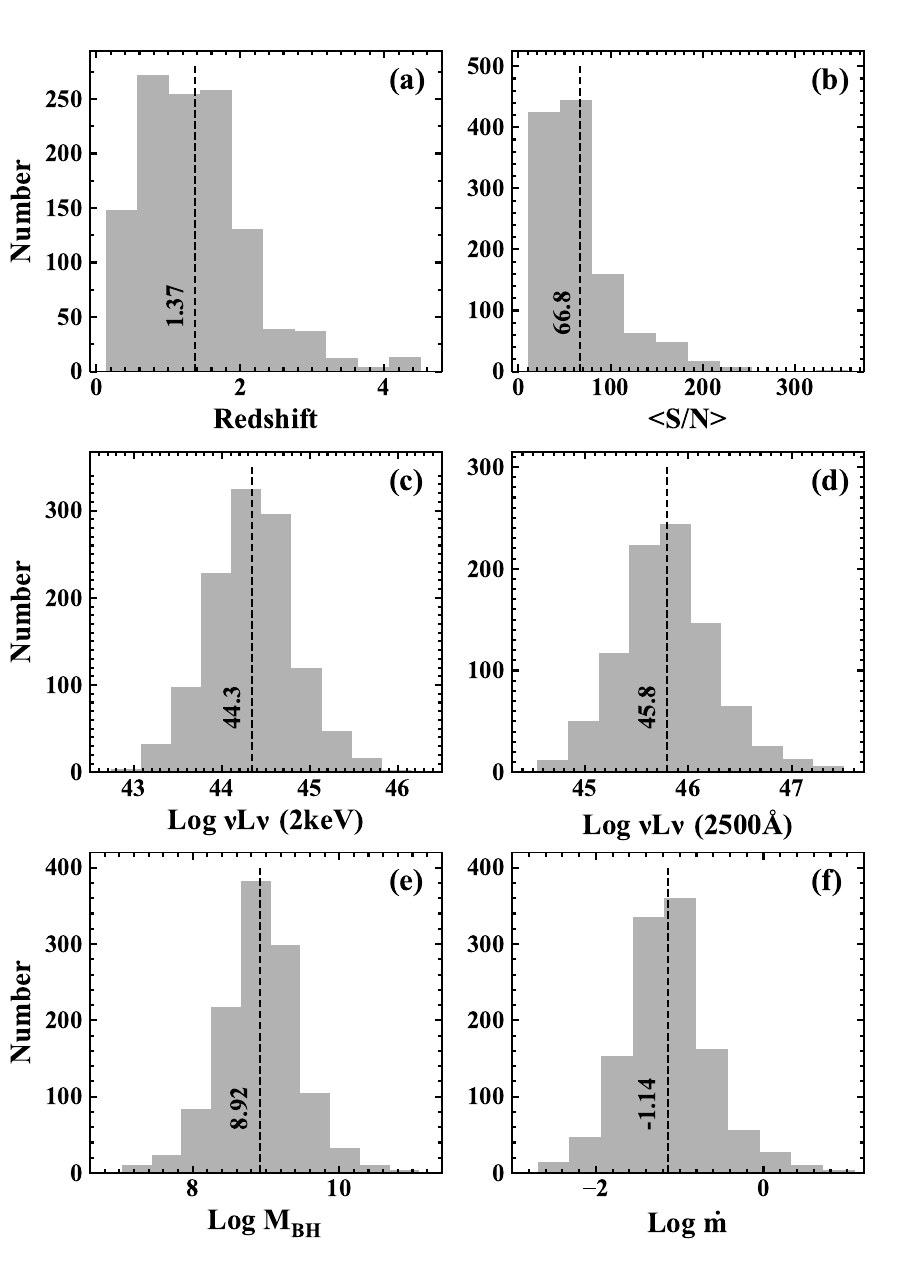}  \\
\caption{Distributions of some basic properties for the main sample, including the redshift (a), mean S/N of the SDSS spectra (b), monochromatic luminosity at 2 keV (c) and 2500 \AA\ (d), black hole mass (e) and mass accretion rate (f). The mean value is indicated by the vertical dash line.}
\label{fig-dist}
\end{figure}

\begin{figure*}
\centering
\begin{tabular}{cc}
\includegraphics[trim=0.3in 0.0in 0.0in 0.0in, clip=1, scale=0.43]{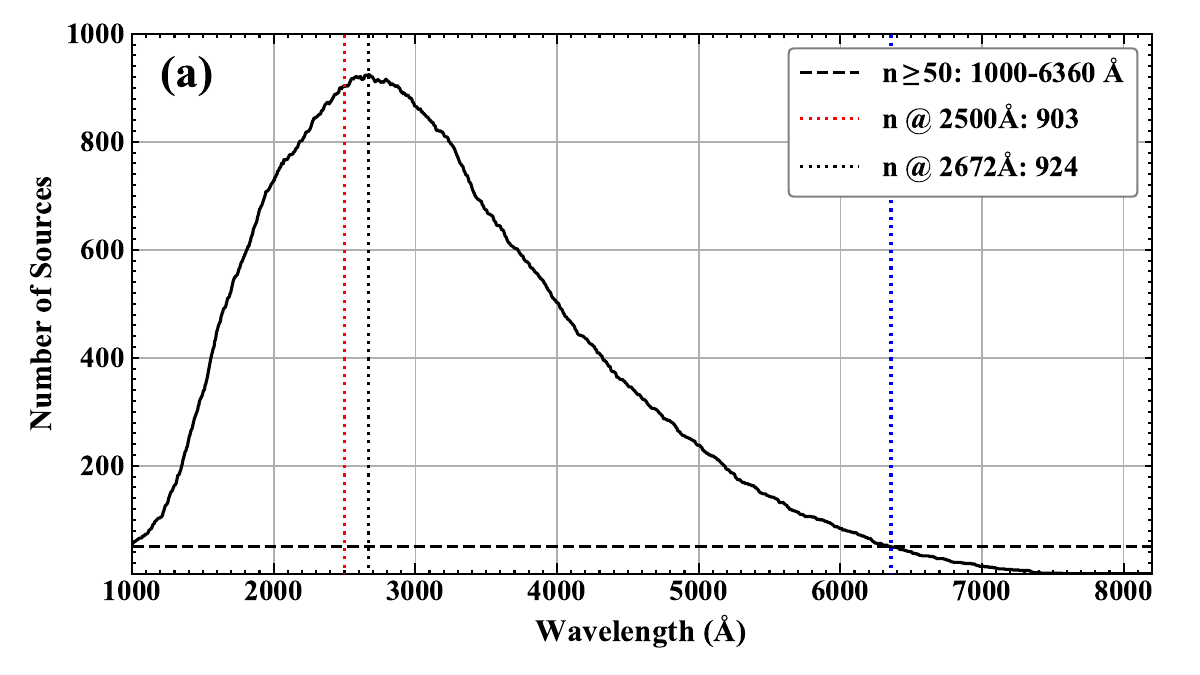}  &
\includegraphics[trim=0.0in 0.0in 0.0in 0.0in, clip=1, scale=0.43]{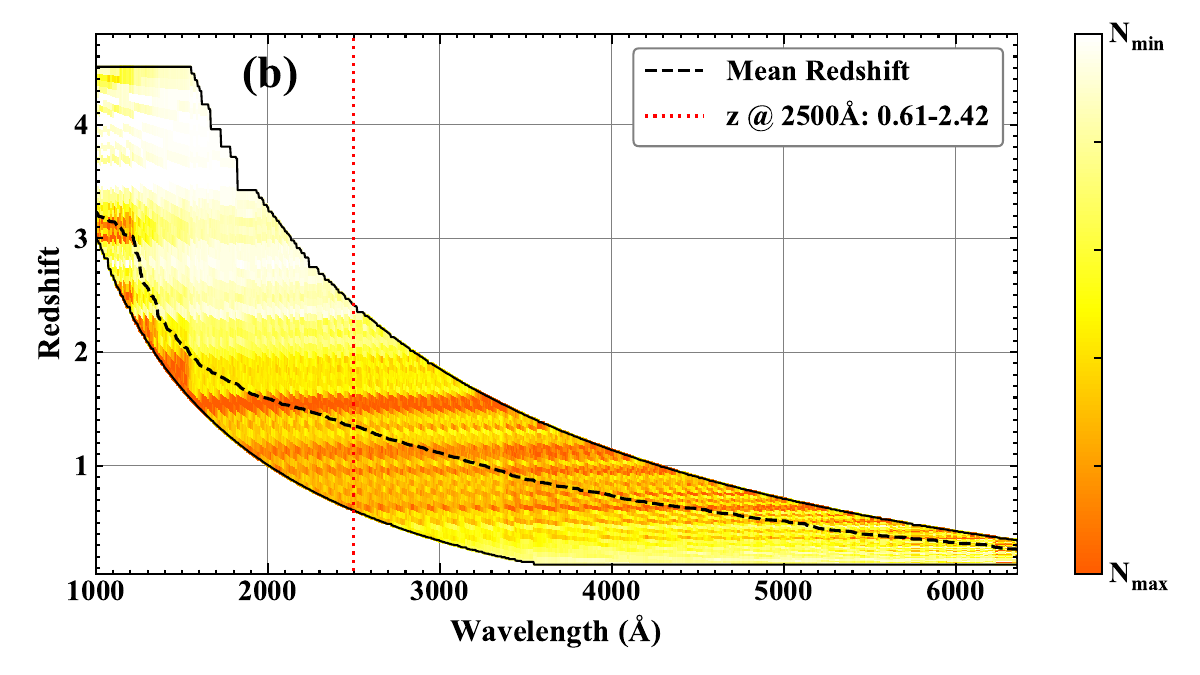}  \\
\includegraphics[trim=0.3in 0.2in 0.0in 0.0in, clip=1, scale=0.43]{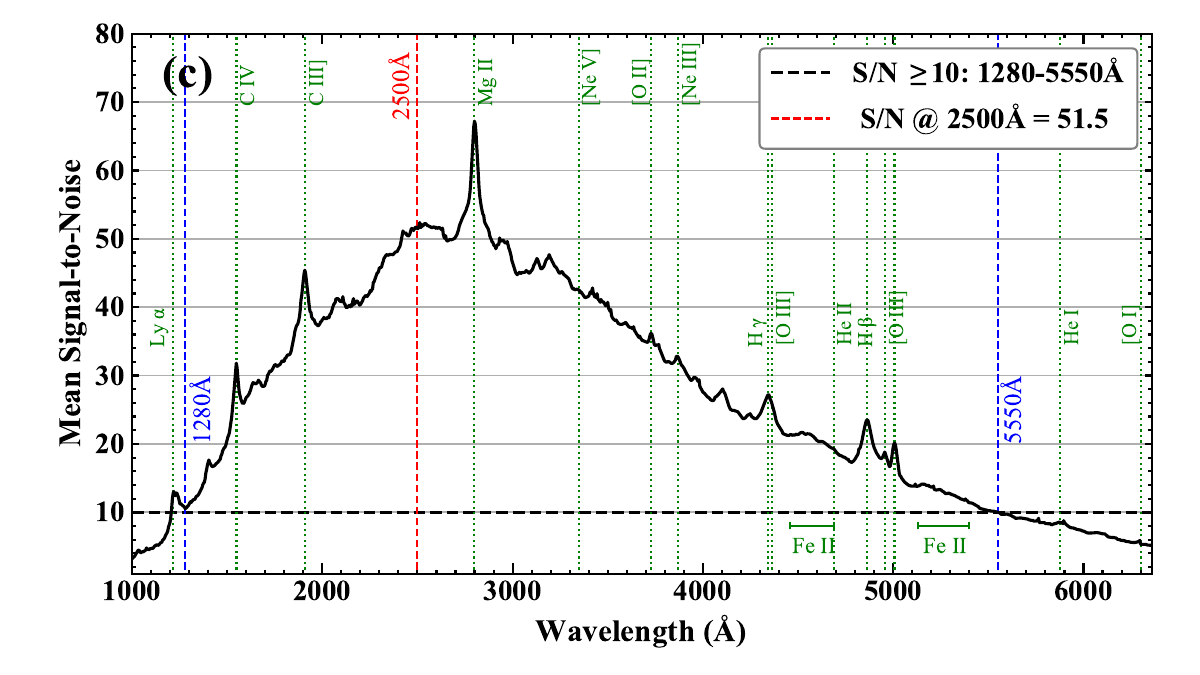}  &
\includegraphics[trim=0.0in 0.2in 0.0in 0.0in, clip=1, scale=0.43]{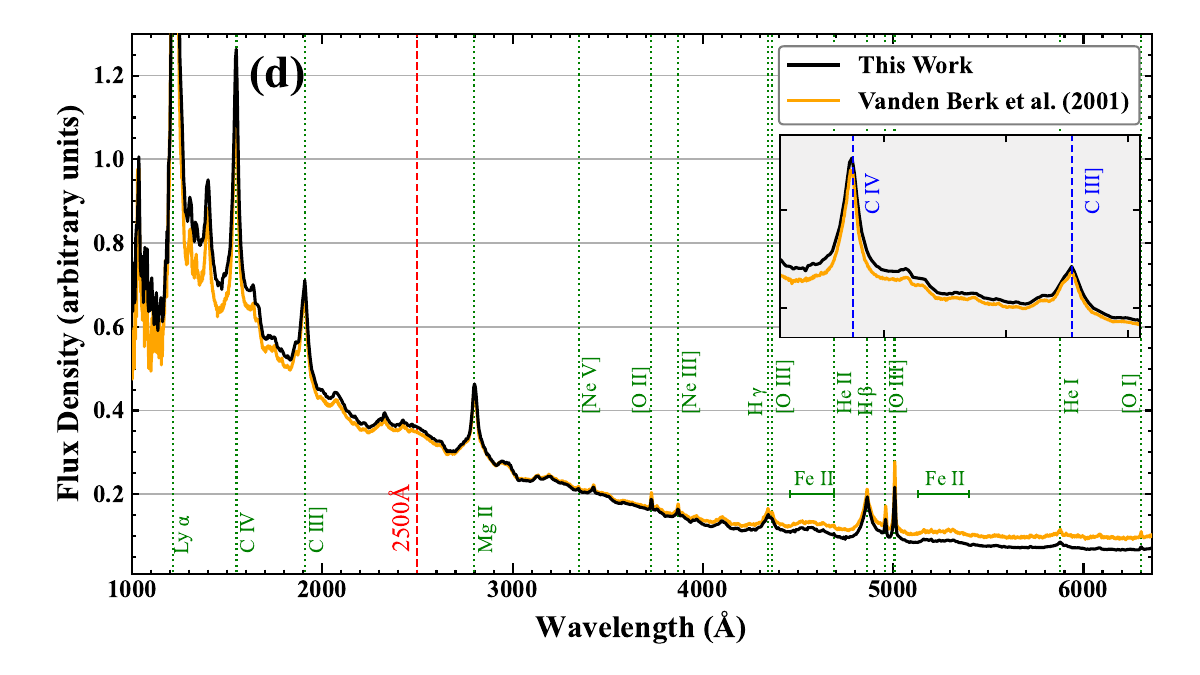}  \\
\end{tabular}
\caption{Panel-a: number of quasars in every wavelength bin whose rest-frame SDSS spectra cover that wavelength. A threshold of $n=50$ is chosen, which restricts the rest-frame wavelength range to 1000 -- 6360 \AA. Panel-b: the redshift distribution of quasars in every wavelength bin. The dash line shows the mean redshift in every wavelength bin. Panel-c: the mean spectral S/N in every wavelength bin. The highest S/N is found in the Mg {\sc ii} line. Panel-d: the composite spectrum\textsuperscript{\ref{spec_dropbox}} for the entire sample (black line), which is slightly steeper than the composite spectrum produced by \citet{Vanden.2001} (orange line).}
\label{fig-wdist}
\end{figure*}

\begin{table}
\centering
   \caption{Range and mean value of some key parameters for the entire sample. $z$ is the redshift. S/N is the signal-to-noise of the SDSS spectra. $L_{\rm 2keV}$ and $L_{\rm 2500\textup{\AA}, phot}$ are the monochromatic luminosities at 2 keV and 2500 \AA. $M_{\rm BH}$ and $\dot{m}$ are the black hole mass and mass accretion rate.}
    \begin{tabular}{lccc}
     \hline
     Parameter & Range & Mean Value & Unit \\
     \hline
     $z$ & 0.13 -- 4.51 & 1.37 & \\
     S/N &  10.7 -- 357.3  & 66.8 & \\
     $L_{\rm 2keV}$ & 42.7 -- 46.2 & 44.3 & erg s$^{-1}$ in log \\
     $L_{\rm 2500\textup{\AA}, phot}$ & 44.5 -- 47.5 & 45.8 & erg s$^{-1}$ in log  \\
     $M_{\rm BH}$ &7.04 -- 11.09 & 8.92 & $M_{\odot}$ in log \\
     $\dot{m}$ & -2.70 -- 1.09 & -1.14 & in log \\
     \hline
     \end{tabular}
\label{tab-dist}
\end{table}

\subsection{Sample Properties and Distributions}
\label{sec-dist}
The redshift distributions of the luminosities at 2500 \AA\ and 2 keV are shown in the right panel of Figure~\ref{fig-2500cross}. These are similar to the parent sample in \citet{Lusso.2020}. Since we applied a S/N cut, some high-redshift sources in the parent sample were excluded due to their low spectral quality, so the redshift range of the current sample is 0.13 -- 4.51 with a mean redshift of 1.37 (see the distributions of various parameters in Table~\ref{tab-dist} and Figure~\ref{fig-dist}a). The mean spectral S/N is 66.8 for the entire sample, confirming that the sample has good SDSS spectral quality. The luminosity at 2 keV covers the range of 42.7 -- 46.2 with a mean value of 44.3 (erg s$^{-1}$ in logarithm), and the luminosity at 2500 \AA\ has a range of 44.5 -- 47.5 with a mean value of 45.8. 

Then the SDSS DR14 catalogue of \citet{Rakshit.2020} is used to extract the black hole masses for the entire sample. These are the fiducial single-epoch virial masses measured from various broad emission lines in the optical/UV band. The mass range covered by our sample is 7.04 -- 11.09 with a mean value of 8.92 in logarithm. Then, given the optical/UV luminosity and the black hole mass, the mass accretion rate through the outer disc ($\dot{m}=\dot{M}/\dot{M}_{\rm Edd}$) can be measured for every source (e.g. \citealt{Davis.2011, Jin.2017b, Jin.2022}). This is done by fitting the SDSS spectrum with the AGN broadband spectral energy distribution (SED) model {\tt optxagnf} (\citealt{Done.2012}), with the black hole mass fixed at the viral mass, spin fixed at zero and $R_{cor}$ fixed at the innermost stable circular orbit i.e. assuming the accretion flow is a standard disc\footnote{A detailed SED study of the sample will be published in another paper.}. Note that $\dot{m}$ is not necessarily equal to the Eddington ratio ($L_{\rm bol}/L_{\rm Edd}$), because the bolometric luminosity ($L_{\rm bol}$) also depends on the radiative efficiency of the accretion flow, which in turn depends on the physical properties of the flow and the black hole spin (e.g. \citealt{Davis.2011, Done.2013, Kubota.2019}). The viewing angle may also affect the observed $L_{\rm bol}$ (e.g. \citealt{Jin.2017b, Jin.2022}). Hence we prefer to use $\dot{m}$ as a more robust indicator of the mass accretion state of an AGN. We find that $\dot{m}$ of the sample covers a range of -2.70 -- 1.09 with a mean value of -1.14 in logarithm (i.e. the mean $\dot{m}$ is 7.2 per cent). 

Besides the overall statistical properties presented above, we also check the sample distribution and quality at various wavelengths. Firstly, we examine the sample size at different wavelengths in the rest-frame, as shown in Figure~\ref{fig-wdist}a. The maximal number (926 quasars) is found at the wavelength of 2672 \AA, which then decreases towards both sides. There are 903 quasars at 2500 \AA. We choose a conservative threshold of $n=50$ to ensure that the correction analysis at each wavelength is statistically robust. This threshold corresponds to a wavelength cut at 6360 \AA, and so we restrict the following analysis to the rest-frame wavelength range of 1000 -- 6360 \AA.

Then, we check the redshift distribution and the mean S/N of different sources in each wavelength bin, as shown Figures~\ref{fig-wdist}b and ~\ref{fig-wdist}c. The sample's redshift range increases naturally towards short wavelengths. At $\lambda \sim$ 1000 \AA, the redshifts are mostly above 3. The redshift range at 2500 \AA\ is 0.61 -- 2.42. For the S/N, the mean value reaches its maximum of 67 at the Mg {\sc ii} line, and decreases towards both sides. The mean S/N at 2500 \AA\ is 51.5. The wavelength range with the mean S/N $\ge$ 10 is 1280 -- 5550 \AA, where the results of correlation analysis should be more reliable.

Finally, we examine the composite optical/UV spectrum for the entire sample. As described in \citet{Vanden.2001}, the composite spectrum is sensitive to the method of assembly. Since the optical/UV underlying continua of quasars generally have a power law shape, we choose the geometric mean algorithm, so that the continuum's shape of the composite spectrum reflects the mean slope of the input sample. To further suppress the bias due to the sample variation at different wavelengths, we divide the sample into three subsamples for three redshift ranges (0.13 -- 0.8, 0.8 -- 1.8, 1.8 -- 4.51), and produce a composite spectrum in each interval. With these three composite spectra, we use the geometric mean to connect different spectra in the two wavebands of 2300 -- 3000 \AA\ and 1470 -- 3000 \AA, and finally derive a total composite spectrum. The absolute flux of the composite spectrum is trivial, so the choice of the above wavelength ranges are arbitrary. The resulting composite spectrum{\footnote{Our new quasar composite spectrum can be downloaded from \url{https:// www.dropbox.com/s/mk3zgwnxe9w66tr/quasar\_compspec\_Jin2023.txt?dl=0}\label{spec_dropbox}}} is shown in Figure~\ref{fig-wdist}d, and is compared to the quasar composite spectrum of \citet{Vanden.2001} which contains 2204 quasars covering the redshift range of 0.044 -- 4.789. We find that these two composite spectra are very similar. The main difference is that the underlying continuum of our new composite spectrum is slightly steeper, which is likely because the parent sample of this study has been filtered more strongly to exclude sources with significant extinction and host galaxy contamination (\citealt{Lusso.2020}).

\begin{table*}
\centering
   \caption{Inter-band Correlation parameters at various wavelengths, including emission lines luminosities and their underlying continual luminosities. $\rho_{\rm p}$ and $\rho_{\rm s}$ are the Pearson and Spearman's correlation coefficients. $\gamma$, $\beta$ and $\delta$ are the slope, intercept and intrinsic dispersion found by the LINMIX regression. $a$ and $b$ are the slope and intercept found by the ODR. $n$ is the number of sources in each subsample. Error bars are for the 1 $\sigma$ confidence range. The correlation plots are shown in Figures~\ref{fig-contcor} and \ref{fig-linecor}.}
    \begin{tabular}{@{}lcccccccc@{}}
     \hline
      & & \multicolumn{2}{c}{Correlation Coeficients} & \multicolumn{3}{c}{LINMIX Parameters} & \multicolumn{2}{c}{ODR Parameters} \\
      & $n$ & $\rho_{\rm p}$ & $\rho_{\rm s}$ & $\gamma$  & $\beta$ & $\delta$ & $a$ & $b$ \\
      \hline
      \multicolumn{8}{l}{$L_{\rm optuv}$ vs. $L_{\rm 2~keV}$} \\
      @1500 \AA & 334 & 0.794 & 0.782 & 0.630 $\pm$ 0.027 & 7.47 $\pm$ 0.85 & 0.225 $\pm$ 0.010 & 0.827 $\pm$ 0.024 & 1.44 $\pm$ 0.77 \\
      @2500 \AA & 903 & 0.786 & 0.768 & 0.643 $\pm$ 0.017 & 6.92 $\pm$ 0.52 & 0.229 $\pm$ 0.006 & 0.759 $\pm$ 0.013 & 3.49 $\pm$ 0.42 \\
      @3500 \AA & 674 & 0.768 & 0.761 & 0.644 $\pm$ 0.021 & 6.86 $\pm$ 0.62 & 0.249 $\pm$ 0.007 & 0.686 $\pm$ 0.019 & 5.71 $\pm$ 0.58 \\
      @4500 \AA & 349 & 0.657 & 0.644 & 0.623 $\pm$ 0.038 & 7.47 $\pm$ 1.15 & 0.271 $\pm$ 0.010 & 0.630 $\pm$ 0.032 & 7.43 $\pm$ 0.97 \\
      @5500 \AA & 143 & 0.549 & 0.539 & 0.605 $\pm$ 0.078 & 7.98 $\pm$ 2.32 & 0.282 $\pm$ 0.017 & 0.474 $\pm$ 0.056 & 12.10 $\pm$ 1.66 \\
      C {\sc iv} Line & 227 & 0.732 & 0.665 & 0.558 $\pm$ 0.034 & 10.68 $\pm$ 1.03 & 0.242 $\pm$ 0.013 & 0.734 $\pm$ 0.030 & 5.47 $\pm$ 0.89 \\
      C {\sc iv} Cont & 229 & 0.833 & 0.803 & 0.650 $\pm$ 0.029 & 6.86 $\pm$ 0.90 & 0.192 $\pm$ 0.011 & 0.680 $\pm$ 0.020 & 5.96 $\pm$ 0.64 \\
      C {\sc iii}] Line & 482 & 0.617 & 0.578 & 0.463 $\pm$ 0.027 & 13.47 $\pm$ 0.79 & 0.277 $\pm$ 0.010 & 0.747 $\pm$ 0.025 & 5.26 $\pm$ 0.73 \\
      C {\sc iii}] Cont & 490 & 0.767 & 0.716 & 0.615 $\pm$ 0.024 & 7.89 $\pm$ 0.73 & 0.224 $\pm$ 0.008 & 0.637 $\pm$ 0.018 & 7.31 $\pm$ 0.55 \\
      Mg {\sc ii} Line & 816 & 0.820 & 0.807 & 0.703 $\pm$ 0.017 & 6.56 $\pm$ 0.49 & 0.208 $\pm$ 0.005 & 0.823 $\pm$ 0.012 & 3.18 $\pm$ 0.36 \\
      Mg {\sc ii} Cont & 816 & 0.769 & 0.754 & 0.636 $\pm$ 0.019 & 7.14 $\pm$ 0.57 & 0.235 $\pm$ 0.006 & 0.703 $\pm$ 0.015 & 5.22 $\pm$ 0.46 \\
      H$\beta$ Line & 198 & 0.725 & 0.694 & 0.656 $\pm$ 0.045 & 7.76 $\pm$ 1.26 & 0.245 $\pm$ 0.013 & 0.789 $\pm$ 0.049 & 4.12 $\pm$ 1.38 \\
      H$\beta$ Cont & 207 & 0.650 & 0.632 & 0.667 $\pm$ 0.054 & 6.17 $\pm$ 1.61 & 0.265 $\pm$ 0.014 & 0.547 $\pm$ 0.039 & 9.91 $\pm$ 1.17 \\\relax
      [O {\sc iii}] $\lambda$5007 Line & 207 & 0.669 & 0.665 & 0.630 $\pm$ 0.049 & 8.83 $\pm$ 1.33 & 0.258 $\pm$ 0.013 & 0.815 $\pm$ 0.045 & 3.73 $\pm$ 1.25 \\\relax
      [O {\sc iii}] $\lambda$5007 Cont & 207 & 0.648 & 0.632 & 0.670 $\pm$ 0.055 & 6.08 $\pm$ 1.63 & 0.266 $\pm$ 0.014 & 0.548 $\pm$ 0.039 & 9.89 $\pm$ 1.18 \\
      \hline
      \multicolumn{8}{l}{$L_{\rm optuv}$ vs. $L_{\rm 2500\text{\AA},phot}$} \\
      @1500 \AA & 334 & 0.983 & 0.981 & 0.975 $\pm$ 0.010 & 0.90 $\pm$ 0.30 & 0.082 $\pm$ 0.003 & 0.994 $\pm$ 0.008 & 0.31 $\pm$ 0.26 \\
      @2500 \AA & 903 & 0.989 & 0.986 & 0.982 $\pm$ 0.005 & 0.53 $\pm$ 0.15 & 0.066 $\pm$ 0.002 & 0.995 $\pm$ 0.004 & 0.12 $\pm$ 0.14 \\
      @3500 \AA & 674 & 0.983 & 0.982 & 0.982 $\pm$ 0.007 & 0.50 $\pm$ 0.21 & 0.083 $\pm$ 0.006 & 0.999 $\pm$ 0.006 & -0.03 $\pm$ 0.19 \\
      @4500 \AA & 349 & 0.960 & 0.959 & 1.016 $\pm$ 0.016 & -0.53 $\pm$ 0.48 & 0.112 $\pm$ 0.004 & 1.034 $\pm$ 0.013 & -1.09 $\pm$ 0.40 \\
      @5500 \AA & 143 & 0.921 & 0.892 & 1.081 $\pm$ 0.039 & -2.51 $\pm$ 1.16 & 0.140 $\pm$ 0.009 & 1.120 $\pm$ 0.026 & -3.62 $\pm$ 0.79 \\
      C {\sc iv} Line & 227 & 0.807 & 0.803 & 0.786 $\pm$ 0.039 & 8.14 $\pm$ 1.14 & 0.280 $\pm$ 0.014 & 1.135 $\pm$ 0.037 & -2.32 $\pm$ 1.10 \\
      C {\sc iv} Cont & 229 & 0.986 & 0.981 & 0.983 $\pm$ 0.010 & 0.63 $\pm$ 0.33 & 0.072 $\pm$ 0.004 & 0.974 $\pm$ 0.009 & 0.93 $\pm$ 0.29 \\
      C {\sc iii}] Line & 482 & 0.818 & 0.807 & 0.745 $\pm$ 0.024 & 9.39 $\pm$ 0.69 & 0.241 $\pm$ 0.008 & 1.039 $\pm$ 0.020 & 0.76 $\pm$ 0.60 \\
      C {\sc iii}] Cont & 490 & 0.984 & 0.973 & 0.950 $\pm$ 0.008 & 1.64 $\pm$ 0.24 & 0.076 $\pm$ 0.003 & 0.962 $\pm$ 0.007 & 1.27 $\pm$ 0.23 \\
      Mg {\sc ii} Line & 816 & 0.937 & 0.925 & 0.958 $\pm$ 0.013 & 3.25 $\pm$ 0.36 & 0.154 $\pm$ 0.004 & 1.059 $\pm$ 0.012 & 0.36 $\pm$ 0.34 \\
      Mg {\sc ii} Cont & 816 & 0.987 & 0.985 & 0.973 $\pm$ 0.005 & 0.79 $\pm$ 0.17 & 0.070 $\pm$ 0.002 & 0.982 $\pm$ 0.005 & 0.52 $\pm$ 0.16 \\
      H$\beta$ Line & 198 & 0.946 & 0.943 & 0.950 $\pm$ 0.022 & 3.25 $\pm$ 0.63 & 0.117 $\pm$ 0.007 & 1.064 $\pm$ 0.023 & 0.05 $\pm$ 0.64 \\
      H$\beta$ Cont & 207 & 0.951 & 0.940 & 1.059 $\pm$ 0.024 & -1.81 $\pm$ 0.72 & 0.118 $\pm$ 0.006 & 1.063 $\pm$ 0.019 & -1.92 $\pm$ 0.57 \\\relax
      [O {\sc iii}] $\lambda$5007 Line & 207 & 0.698 & 0.672 & 0.716 $\pm$ 0.052 & 10.18 $\pm$ 1.42 & 0.274 $\pm$ 0.014 & 1.076 $\pm$ 0.061 & 0.26 $\pm$ 1.68 \\\relax
      [O {\sc iii}] $\lambda$5007 Cont & 207 & 0.949 & 0.937 & 1.064 $\pm$ 0.025 & -1.96 $\pm$ 0.75 & 0.121 $\pm$ 0.006 & 1.066 $\pm$ 0.019 & -2.00 $\pm$ 0.58 \\
     \hline
     \end{tabular}
\label{tab-linecor}
\end{table*}

\section{The Direct Correlation Spectrum}
\label{sec-optuvx}
The correlation between the optical/UV and X-ray luminosities of quasars has been extensively studied, generally based on their luminosity at 2500 \AA. The method of direct optical to X-ray correlation spectrum (OXCS, \citealt{Jin.2012b}) offers a new model-independent way to investigate the correlation with X-rays at different wavelengths in the optical band. By applying this method to a sample of 51 unobscured AGN with high-quality multi-wavelength data, \citet{Jin.2012b} found that the optical continuum is generally well correlated with the X-ray emission, and some principal emission lines (e.g. the broad H$\beta$, [OIII] $\lambda$5007) correlate better with the X-rays than with the continuum. In this section we extend the OXCS method by applying it to a much larger sample and over a wider wavelength range, including both optical and UV, and then replace the X-ray luminosity with the UV luminosity to explore, for the first time, the optical/UV self-correlation.

\subsection{Optical/UV and X-ray Correlation Spectrum (OUXCS)}
\label{sec-oxcs}
\subsubsection{The First Type: OUXCS-1}
\begin{figure}
\centering
\begin{tabular}{c}
\includegraphics[trim=0.5in 0.3in 0.0in 0.0in, clip=1, scale=0.46]{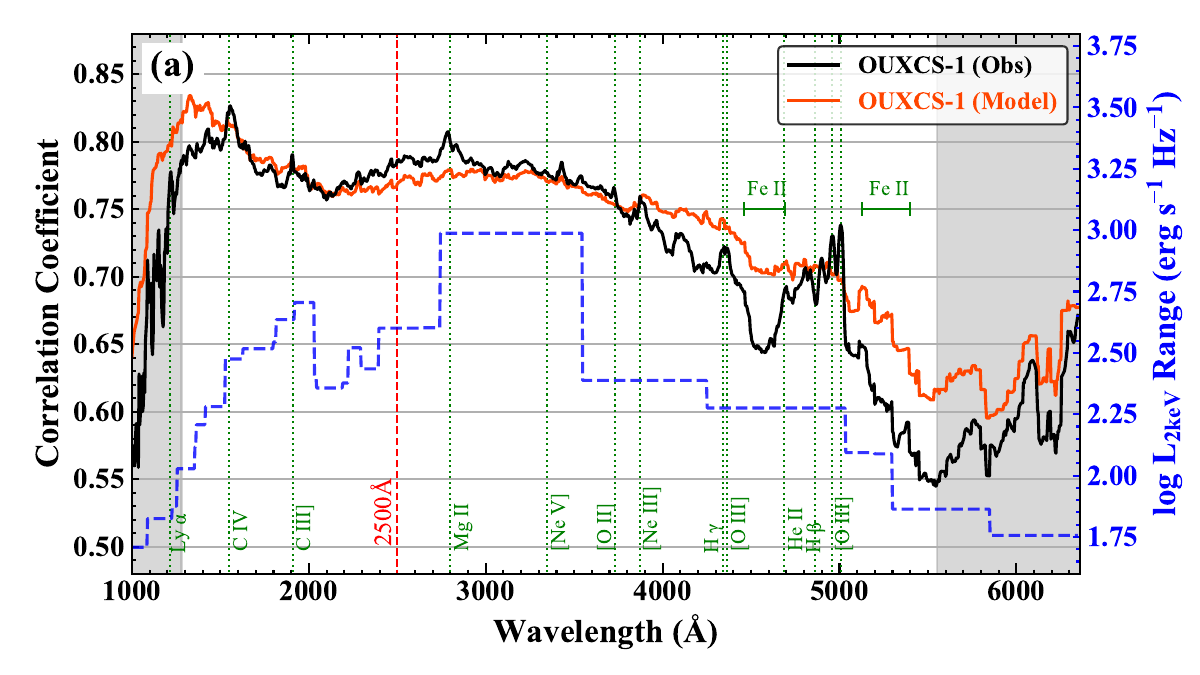}  \\
\includegraphics[trim=0.5in 0.3in 0.0in 0.0in, clip=1, scale=0.46]{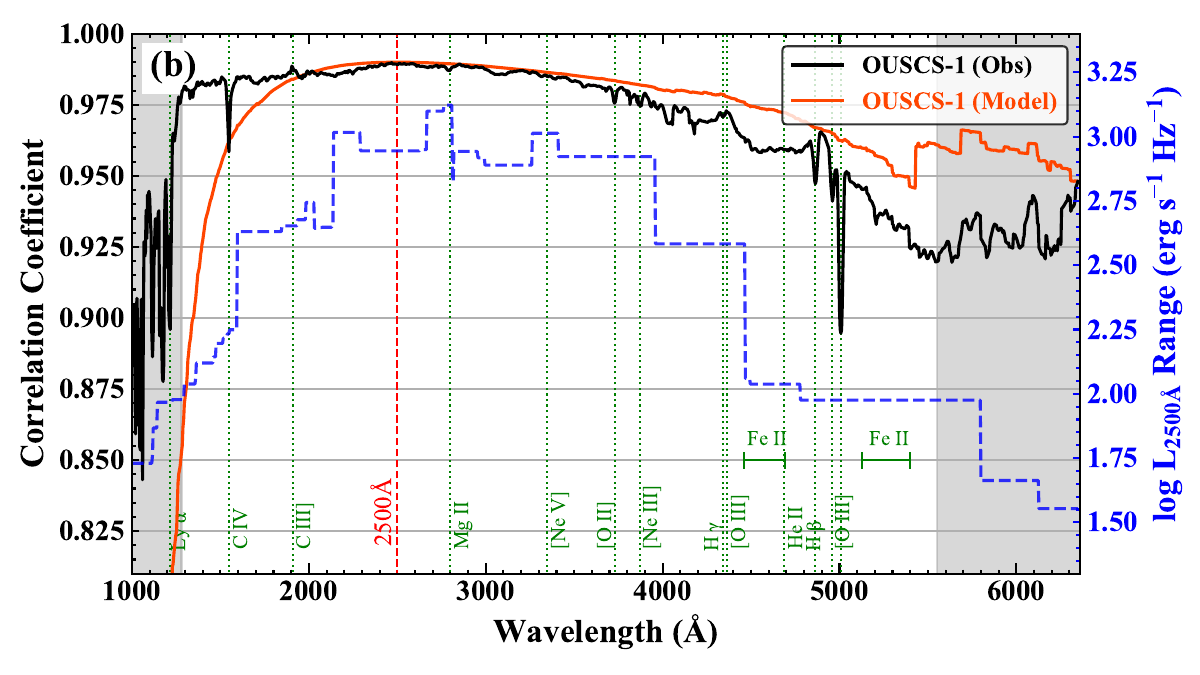}  \\
\end{tabular}
\caption{Panel-a: the first type of OUXCS (i.e. OUXCS-1). The black solid line shows the observed OUXCS-1. The red solid line shows the OUXCS-1 derived from the disc model. The blue dash line shows the range of 2 keV luminosity for the subsample at each wavelength. Panel-b: similar to Panel-a, but for the first type of OUSCS (i.e. OUSCS-1). The blue dash line shows the range of 2500 \AA\ luminosity. The grey regions on both sides are the low S/N regions (mean S/N < 10).}
\label{fig-oxcs}
\end{figure}

Following the OXCS method we cross-correlate the monochromatic luminosity at each wavelength within 1000 -- 6300 \AA\ with the 2 keV luminosity ($L_{\rm 2~keV}$) for the subsample available at that wavelength, and then plot the Pearson correlation coefficient against the wavelength (hereafter: the first type of optical/UV and X-ray correlation spectrum, OUXCS-1). The results are shown in Figure~\ref{fig-oxcs}a (black solid line). The data below 1280 \AA\ and above 5550 \AA\ are affected by the low spectral quality (see Figure~\ref{fig-wdist}c), so we only consider the results in the range of 1280 -- 5550 \AA\ to be reliable.

Firstly, OUXCS-1 confirms the results reported previously by \citet{Jin.2012b}, including the strong optical continuum's correlation and stronger line correlations for H$\beta$ and [O {\sc iii}]$\lambda$5007. Secondly, OUXCS-1 shows that $L_{\rm 2~keV}$ correlates more strongly with the UV continuum below 3500 \AA\ than with the optical range. In addition, correlations with emission lines such as Mg {\sc ii}, C {\sc iii}] and C {\sc iv} are stronger than with their local continua.

Figure~\ref{fig-wdist}b and Figure~\ref{eq-luvx}b shows that subsamples at different wavelengths span different redshift and luminosity ranges, thus the general shape of the observed OUXCS-1 is mainly driven by the accretion disc continuum properties of the subsample at different wavelengths. To quantify this effect, we considered the  best-fit disc model of {\tt optxagnf} for every source (see Section~\ref{sec-dist}) and recalculate OUXCS-1 directly from the model. The result is shown in Figure~\ref{fig-oxcs}a (red solid line). Indeed, we find that the shape of the model's OUXCS-1 follows the observed OUXCS-1 well, including the drop in correlation at longer wavelengths. The blue line (right hand axis) show the range of X-ray luminosity spanned by the sample at each wavelength. A larger span in luminosity gives a better correlation as the trend is dominated by the intrinsic spectral changes rather than the dispersion (see Figure~\ref{fig-contcor}).
Therefore, the general shape of the observed OUXCS-1 is mainly due to the properties of the subsample at different wavelengths.

There are clearly some correlated emission line features in the observed OUXCS-1, and differences also exist in the correlated continuum below 1500 \AA\ and above 4000 \AA, thus a difference between the model and the observed OUXCS-1 general shape in these ranges could be due to additional spectral features (e.g. emission lines, Fe{\sc ii} complex) with respect to the accretion disc continuum.

\begin{figure}
\centering
\begin{tabular}{c}
\includegraphics[trim=0.2in 0.2in 0.0in 0.0in, clip=1, scale=0.42]{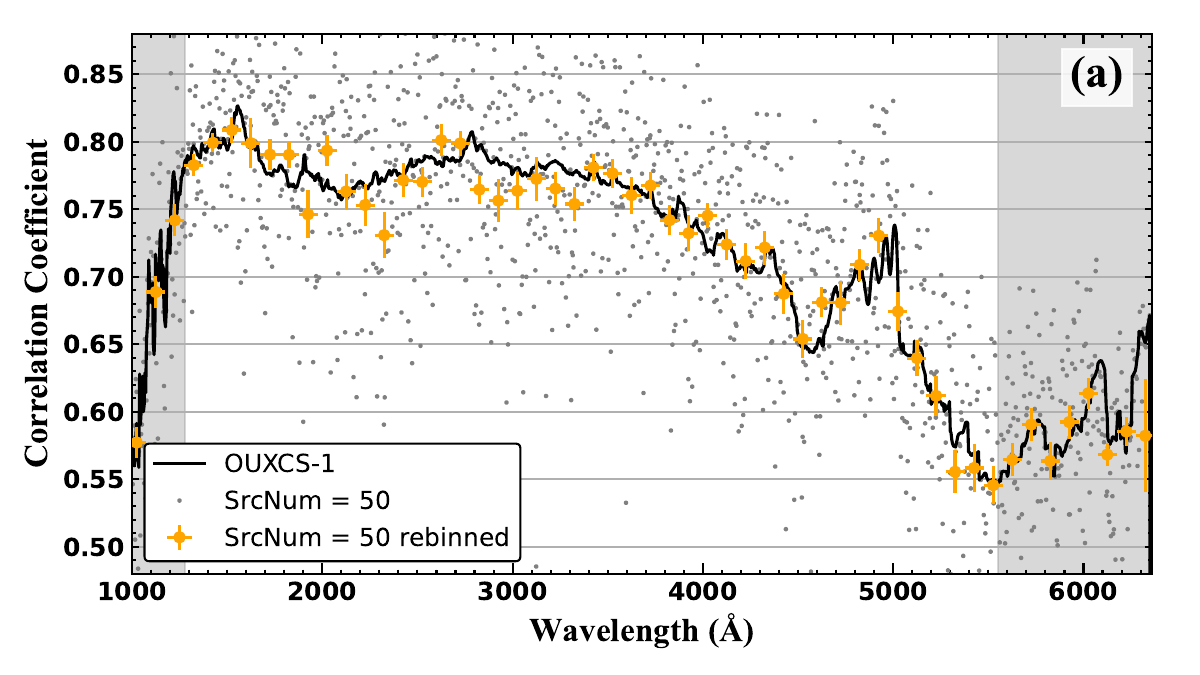}  \\
\includegraphics[trim=0.2in 0.3in 0.0in 0.0in, clip=1, scale=0.42]{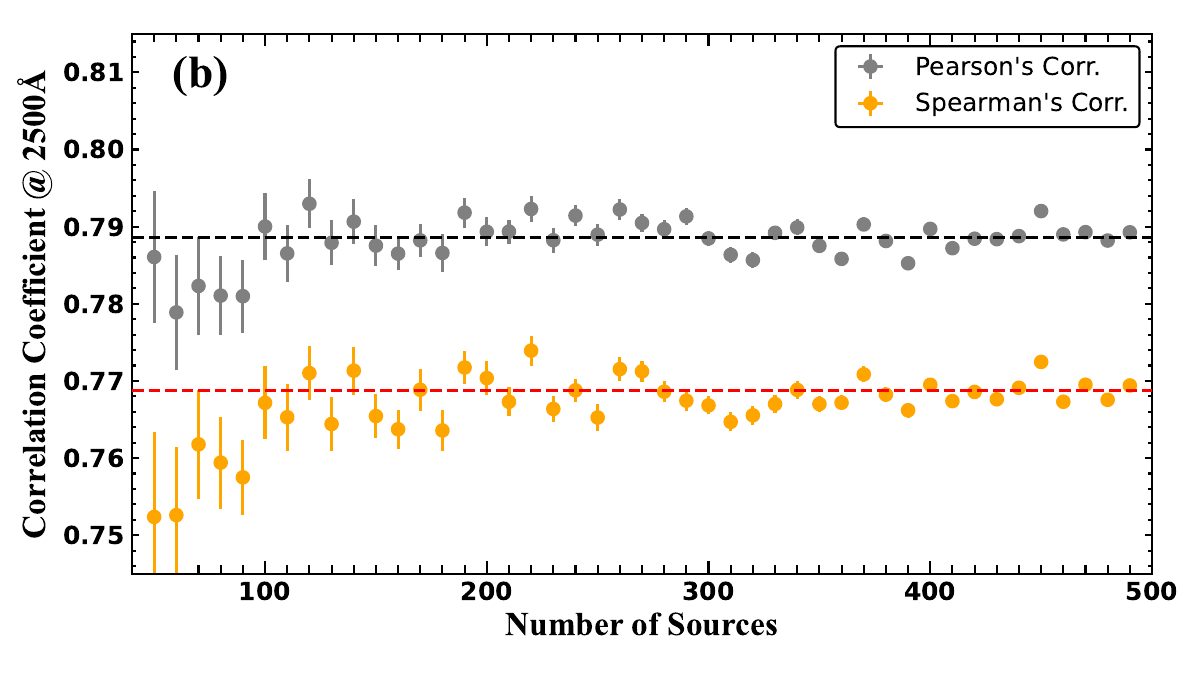}  \\
\end{tabular}
\caption{Examining the effect of sample size in OUXCS-1. Panel-a: randomly selecting 50 sources from the sample at each wavelength and calculate the correlation coefficient (grey points), then rebining these points with 50 points per bin (orange points). Panel-b: at the wavelength of 2500 \AA, randomly selecting sub-samples of different sizes and examining the correlation coefficients. These results indicate that sample size does not affect the OUXCS. The grey regions on both sides are the low S/N regions (mean S/N < 10).}
\label{fig-oxcs-simu}
\end{figure}

We then examine the possible presence of biases driven by subsample's size and luminosity range, separately. To assess the bias of subsample size, we first randomly pick 50 sources from the subsample at each wavelength, and then calculate the correlation coefficient again. In this case, the sample size is wavelength invariant. These random objects are shown with grey points in Figure~\ref{fig-oxcs-simu}a. For visualisation purposes, we rebin the data by using 50 data points per interval (orange points). The rebinned correlation curve is consistent with the original OUXCS-1. We then randomly select subsamples with sizes ranging from 50 to 500 at the wavelength of 2500 \AA\ and calculate their correlation coefficients. We find that the correlation coefficient does not change significantly with the sample size (see Figure~\ref{fig-oxcs-simu}b). Therefore, we conclude that the change of subsample size has a negligible effect on the shape of OUXCS-1.

The change of subsample's luminosity range (i.e. the difference between maximum and minimum luminosity of the sub-sample at each wavelength) may also affect the correlation. The blue dash line in Figure~\ref{fig-oxcs}a shows the range of 2 keV luminosity of the subsample at each wavelength. It peaks at $\sim$ 3000 \AA\ and decreases towards both sides, which is similar to the general shape of OUXCS-1. To understand this effect, we plot the correlations at a series of wavelengths in Figure~\ref{fig-contcor}. It shows that the intrinsic dispersion is similar at these wavelengths (also see Table~\ref{tab-linecor}), and the main difference is the optical/UV luminosity range. It also shows that as the optical/UV luminosity range decreases, the intrinsic dispersion impacts the correlation more strongly. This can be demonstrated in a more quantitative way with the sample of 903 quasars at 2500 \AA. The mean 2500 \AA\ luminosity of this subsample is 30.7 in the units of logarithmic erg s$^{-1}$ Hz$^{-1}$. Then we took the subsample within the luminosity range of $30.7 \pm x$ and calculated the Pearson's correlation coefficient $\rho_{\rm p}$. We found that as $x$ decreases from 1.5 dex to 1.0, 0.75, 0.5, 0.25, 0.15 dex, $\rho_{\rm p}$ decreases from 0.78 to 0.74, 0.70, 0.60, 0.31, 0.17.
Thus we confirm that the shape of OUXCS-1 is affected by the subsample's luminosity range at different wavelengths. However, we also notice that below $\sim$ 2000 \AA, the shape of OUXCS-1 is different from the change of luminosity range, implying that not all curvatures seen in OUXCS-1 can be explained by the change of luminosity range.

\begin{figure*}
\centering
\begin{tabular}{cc}
\includegraphics[trim=0.1in 0.0in 0.0in 0.0in, clip=1, scale=0.42]{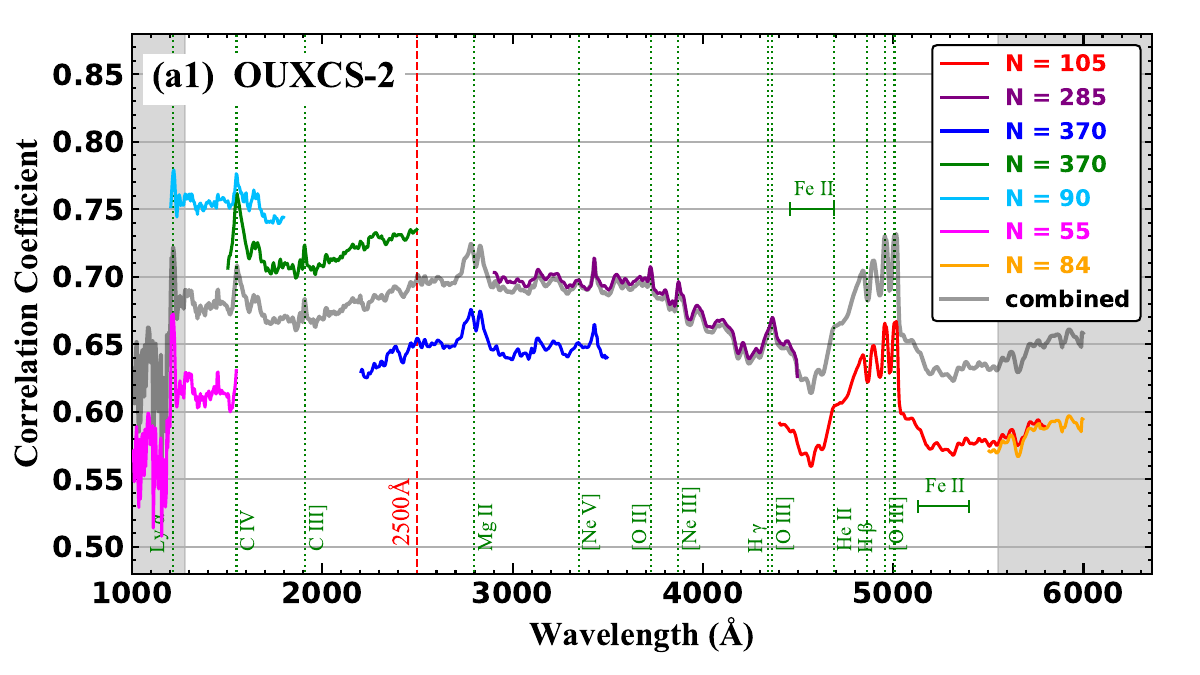}  &
\includegraphics[trim=0.1in 0.0in 0.0in 0.0in, clip=1, scale=0.42]{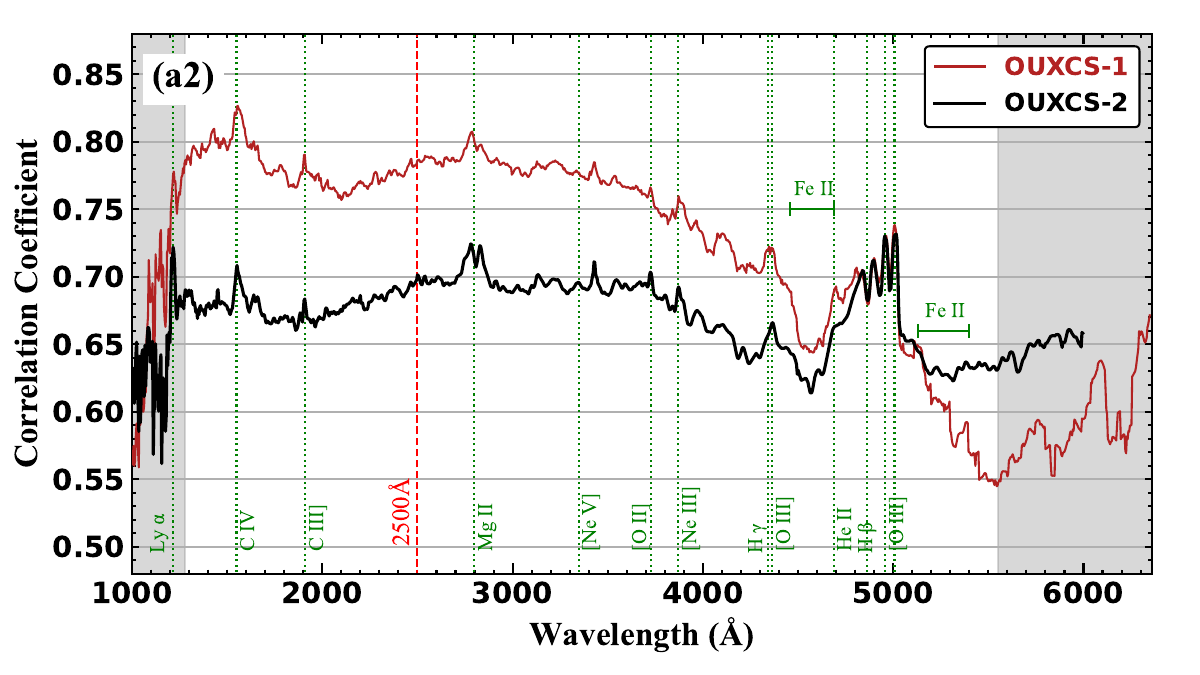}  \\
\includegraphics[trim=0.4in 0.2in 0.0in 0.0in, clip=1, scale=0.42]{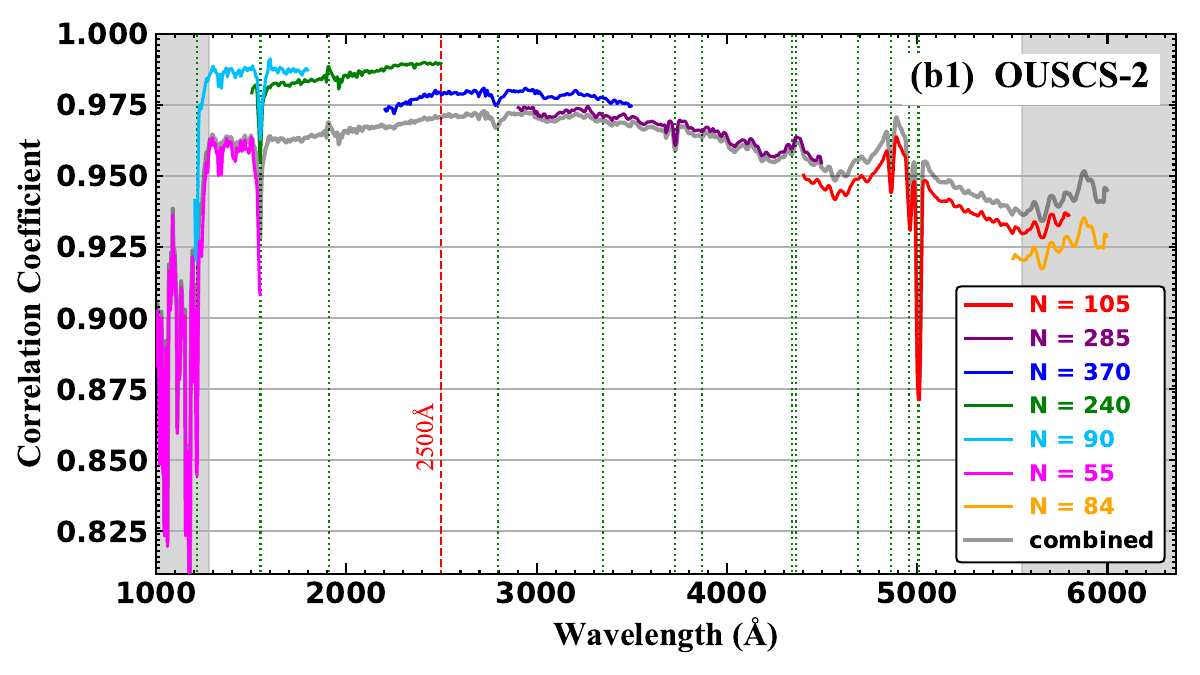}  &
\includegraphics[trim=0.4in 0.2in 0.0in 0.0in, clip=1, scale=0.42]{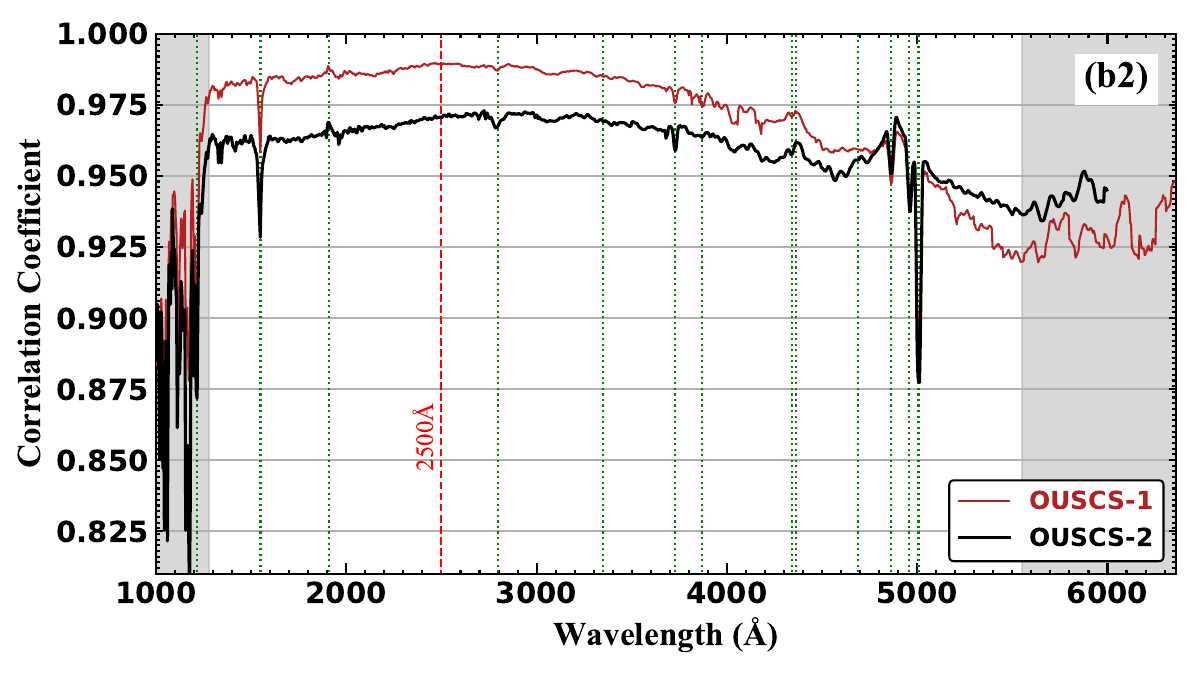}  \\
\end{tabular}
\caption{Panel-a1: the OUXCS at different waveband, shown by different colours, each of them is based on the same subsample. The grey line shows the combined OUXC, i.e. OUXCS-2,  renormalized to OUXCS-1 within 4900 -- 5100 \AA. The number of sources in each subsample is indicated in the legend. Panel-a2: comparison between OUXCS-1 (dark red line) and OUXCS-2 (black line). Panel-b1: similar to Panel-a1, but for OUSCS-2. Panel-b2: comparison between OUSCS-1 (dark red line) and OUSCS-2 (black line). The grey regions on both sides are the low S/N regions (mean S/N < 10).}
\label{fig-oxcs-2}
\end{figure*}

\subsubsection{The Second Type: OUXCS-2}
To minimise the effect of luminosity range, we develop another method to construct OUXCS for a wide wavelength range. Firstly, we divide the entire sample into seven subsamples covering seven different wavelength ranges, including 1000 -- 1550 \AA, 1200 -- 1800 \AA, 1500 -- 2500 \AA, 2200 -- 3500 \AA, 2900 -- 4500 \AA, 4400 -- 5800 \AA\ and 5500 -- 6000 \AA. Then we calculate the OUXCS for each one of them. The seven OUXCS are shown by different colours in Figure~\ref{fig-oxcs-2} Panel-a1. In this case, each OUXCS is built from the same subsample, so its shape is not affected by the change of luminosity range, only the normalization is affected.
Then we renormalize the OUXCS of the subsample at $\sim$ 5000 \AA\ (red line) to match OUXCS-1 within 4900 -- 5100 \AA, and then renormalize the rest of the OUXCS one by one so that they all join together smoothly. We take the average value within their overlapping wavelength ranges. This jointed OUXCS is called the second type of OUXCS (hereafter: OUXCS-2), as shown by the grey line in Figure~\ref{fig-oxcs-2} Panel-a1.

The selection of wavelength range is mainly based on the range's width and subsample's size. The larger the wavelength range, the fewer sources within that range, and the representativeness of correlation is lower. Meanwhile, it is also necessary to have some overlap between different wavelength ranges, so as to evaluate the difference in correlation between different subsamples in the overlap region. The specific choice of wavelength range has a certain degree of arbitrariness, but because the correlation spectra of different subsamples in the overlap regions are basically consistent, so choosing different wavelength ranges should not bring significant differences to the general shape of OUXCS-2. Thus we can conclude that the effect of luminosity range is minimized in OUXCS-2.

Figure~\ref{fig-oxcs-2} Panel-a2 compares these two types of OUXCS. OUXCS-2 appears much flatter than OUXCS-1, which is because the bias in luminosity range is minimized in OUXCS-2, and both confirm similar good correlations across 1280 -- 5550 \AA. OUXCS-1 and OUXCS-2 also show similar line correlations, including a stronger correlation in the broad emission line component of both Mg {\sc ii} and H$\beta$, but weaker in the narrow component. The correlations at C {\sc iv}, C {\sc iii}] and [O {\sc iii}]$\lambda\lambda$4959/5007 are also enhanced significantly. Interestingly, there is some evidence that the Fe {\sc ii} complex before on both sides of H$\beta$ seems to behave the opposite with respect to the other emissions lines. In addition, both OUXCS display stronger correlation with the UV continuum than with the optical, with a correlation parameter that peaks at 2500 -- 3800 \AA. The statistical error on the OUXCS is $\sim$ 0.01, so the difference between optical and UV in OUXCS-2 is statistically significant.

\subsection{Optical/UV Self-Correlation Spectrum (OUSCS)}
\label{sec-ouvcs}
While the 2 keV luminosity may physically represent the intensity of the corona emission, including both the hot corona and probably some fraction of the warm corona, the 2500 \AA\ luminosity could be dominated by emission from the disc itself (although an extended warm corona may also contribute significantly to the UV emission for some AGN, see \citealt{Gardner.2017}). Thus, by taking the luminosity at 2500 \AA, instead of 2 keV as the correlation variable, we can examine the self-correlation of the disc emission in the optical/UV, as well as any correlations for various emission lines with the disc continuum.

The photometric luminosity at 2500 \AA\ ($L_{\rm 2500\textup{\AA},phot}$) is used because it is available for the entire sample \citep[as listed in the catalogue published by][]{Lusso.2020}. This new version of correlation spectrum is named the optical/UV self-correlation spectrum (OUSCS). Likewise, we first create the first type of OUSCS, namely OUSCS-1, as shown in Figure~\ref{fig-oxcs}b (black solid line). The self-correlation with $L_{\rm 2500\textup{\AA},phot}$ is approaching unity across the entire waveband, and is much stronger than the correlations with $L_{\rm 2 keV}$, and the self-correlation is also stronger in the UV band than optical band.

There are many absorption features in OUSCS-1, indicating that the correlation of the emission lines with the UV luminosity is weaker than with their underlying continua. To examine the effect of sample bias, we also calculate the OUSCS-1 from the disc model, as shown by the red solid line in Figure~\ref{fig-oxcs}b. The comparison between the observed and model OUSCS-1 suggests that the overall shape is driven by the disc emission, whilst the small-scale residual fluctuations in the observed OUSCS-1 have a different origin than the disc continuum (e.g. emission lines). The observed OUSCS-1 remains flat below 2000 \AA, which is mostly because the input disc model does not provide a good match of the observed spectra in the UV band. This finding is consistent with the ones published by \citet{Mitchell.2023}. By analysing a sample of $\simeq$700 SDSS AGN, they found that the data do not match the predictions made by any current accretion flow model. Specifically, they observed that either the disc is completely covered by a warm Comptonisation layer, whose properties change with accretion rate, or the accretion flow structure is different to that of the standard disc models (see their Figure 10 and discussion in their Section~5).

The gradual decrease of the observed OUSCS-1 towards longer wavelengths also seems to follow the decrease of the luminosity range (blue dash line), so we create OUSCS within the aforementioned seven different wavebands, each of which is based on the same subsample (see Figure~\ref{fig-oxcs-2} Panel-b1). These individual short OUSCS are then joined together, and renormalized to OUSCS-1 within 4900 -- 5100 \AA, to create the second type of OUSCS, namely OUSCS-2.

Figure~\ref{fig-oxcs-2} Panel-b2 compares the two types of OUSCS. Likewise, OUSCS-2 is flatter than OUSCS-1 after suppressing the bias of luminosity range. The correlations remain good across the entire optical/UV band, and the UV continuum show slightly better correlations than optical. The statistical error of OUSCS is $\sim$ 0.001, so the difference between optical and UV in OUSCS-2 is also statistically significant. The peak correlation lies within 2500 -- 3500 \AA. In addition, OUSCS-2 also shows various absorption-like features. Similar to OUXCS-2, the broad component of H$\beta$ shows stronger correlation, while the correlation of the narrow component is weaker. However, C {\sc iv}, C {\sc iii}], Mg {\sc ii} and [O {\sc iii}]$\lambda\lambda$4959/5007 all appear absorption-like, suggesting that their correlations with $L_{\rm 2500\textup{\AA},phot}$ are weaker than with their underlying continua. The absorption-like features at C {\sc iv} and [O {\sc iii}] are particularly strong. This is probably due to their higher ionization energies ($>$ 50 eV), which also leads to their higher correlations with the 2 keV luminosity in OUXCS. We discuss these in more detail in Section~\ref{sec-linecor}.

\begin{figure}
\centering
\begin{tabular}{c}
\includegraphics[trim=0.2in 0.5in 0.0in 0.0in, clip=1, scale=0.42]{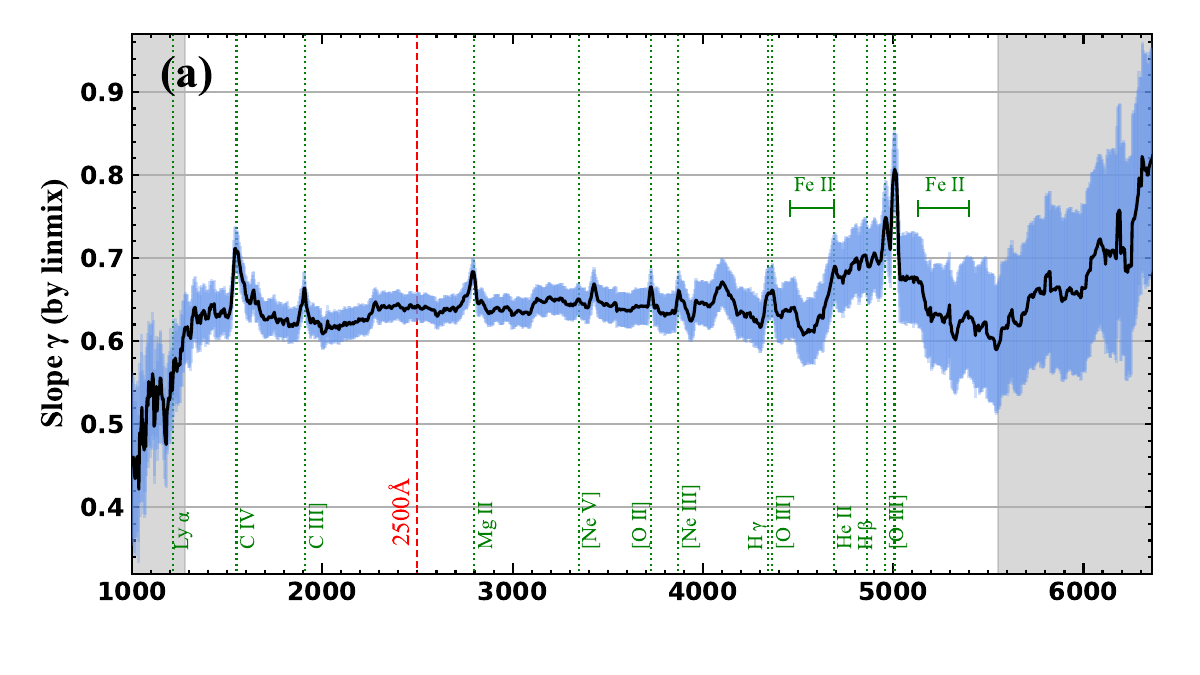}  \\
\includegraphics[trim=0.2in 0.5in 0.0in 0.0in, clip=1, scale=0.42]{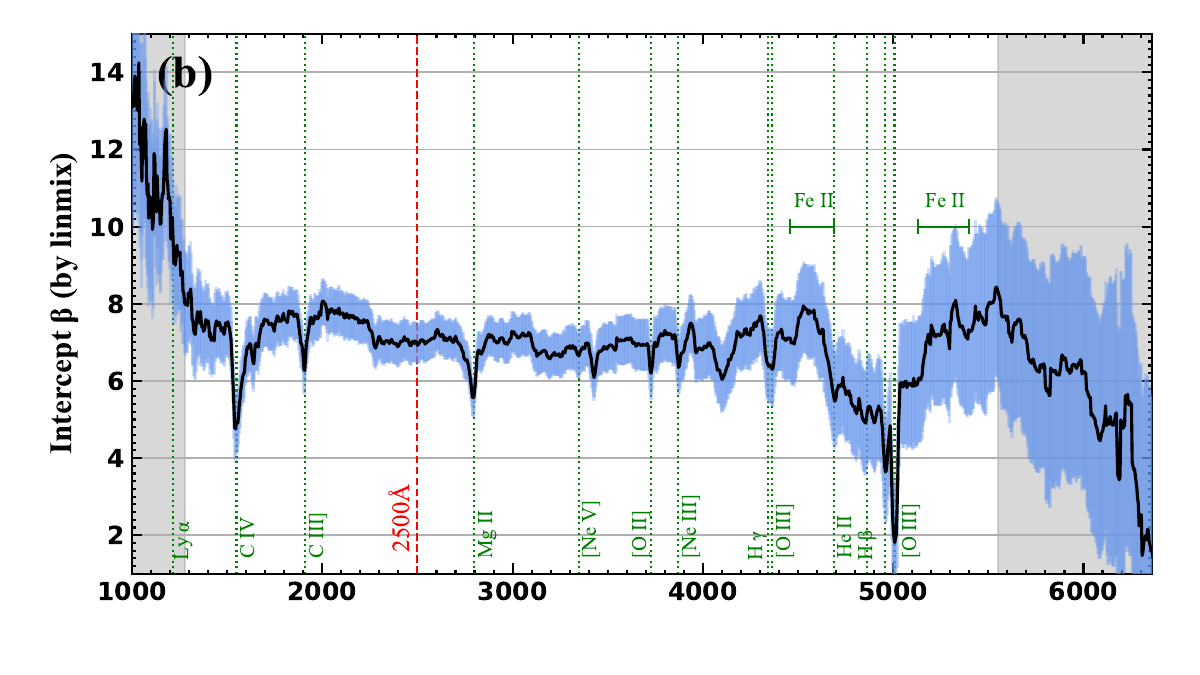}  \\
\includegraphics[trim=0.2in 0.2in 0.0in 0.0in, clip=1, scale=0.42]{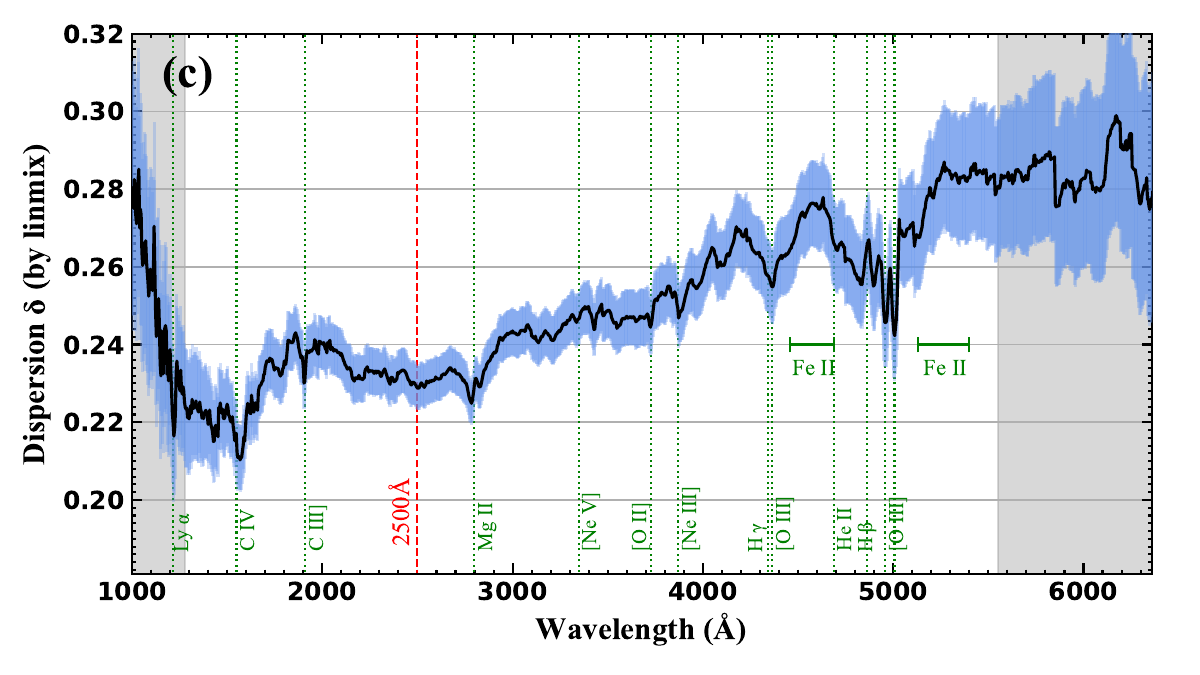}  \\
\end{tabular}
\caption{Wavelength dependence of the optical/UV to X-ray (2 keV) regression parameters, including the slope (Panel-a), intercept (Panel-b) and dispersion (Panel-c) as found by the LINMIX algorithm. The light blue region shows the $\pm~1~\sigma$ uncertainty range at every wavelength. The grey regions at both sides are the low S/N regions (mean S/N < 10).}
\label{fig-linmix}
\end{figure}

\section{Wavelength Dependences of the Inter-band Regression Parameters}
\label{sec-wdep}
Since OUXCS shows consistent strong inter-band correlations between optical/UV and X-ray, it is instructive to derive the regression parameters at every wavelength and to examine their wavelength dependencies. For ease of comparison with previous studies (e.g. \citealt{Lusso.2016}), we use the following equation and then apply the same LINMIX algorithm to perform a regression analysis at each wavelength, 
\begin{equation}
\label{eq-luvx}
{\rm log}(L_{\rm 2~keV})=\gamma\ {\rm log}(L_{\rm optuv}) + \beta
\end{equation}
where $L_{\rm 2~keV}$ and $L_{\rm optuv}$ are the X-ray and optical/UV luminosities in the units of erg s$^{-1}$ Hz$^{-1}$. The regression parameters include the slope $\gamma$ and intercept $\beta$, as well as the intrinsic (random) scatter $\delta$ around the regression.

Figure~\ref{fig-linmix} plots these parameters as a function of wavelength. Firstly, for the direct comparison with previous studies, we derive the regression parameters at 2500 \AA\ with $\gamma=0.643\pm0.017$, $\beta=6.92\pm0.52$ and $\delta=0.229\pm0.006$. These results are statistically consistent with previous studies using the photometric luminosity at 2500 \AA\ (e.g. \citealt{Lusso.2016, Lusso.2020}). Secondly, we find that $\gamma$ is 0.6 -- 0.7 and $\beta$ is 6 -- 8 for the entire continuum within 1250 -- 5550 \AA\ where the sample-averaged S/N is $> 10$, suggesting that the slope and intercept do not have strong wavelength dependences. Thirdly, we find that the intrinsic scatter at 2500 \AA\ is small compared with the other wavelengths. There results suggest that 2500 \AA\ is indeed a good, representative choice that can be used to study the optical/UV and X-ray luminosity correlations.

However, Figure~\ref{fig-linmix} also shows that the regression parameters are sensitive to the presence of emission lines (e.g. C {\sc iv}, C {\sc iii}], Mg {\sc ii}, H$\beta$ and [O {\sc iii}]$\lambda\lambda$4959/5007). Where a spectral line is present the slope and the intercept change significantly. This result suggests  that, compared with the optical/UV continuum, the emission lines have different correlations with the X-rays. Furthermore, we also find that different velocity components of a single emission line can exhibit a different correlation. For example, there is a significant blue wing in the composite C {\sc iv} and C {\sc iii}] line profiles (see Figure~\ref{fig-wdist}d and Section~\ref{sec-compare-method}), but this wing component does not emerge in Figure~\ref{fig-linmix}. This is consistent with the recently reported lack of a correlation between the velocity of these lines and the X-ray luminosity (see Fig.13 in \citealt{Lusso.2021}).

\begin{figure}
\centering
\begin{tabular}{c}
\includegraphics[trim=0.1in 0.2in 0.0in 0.0in, clip=1, scale=0.55]{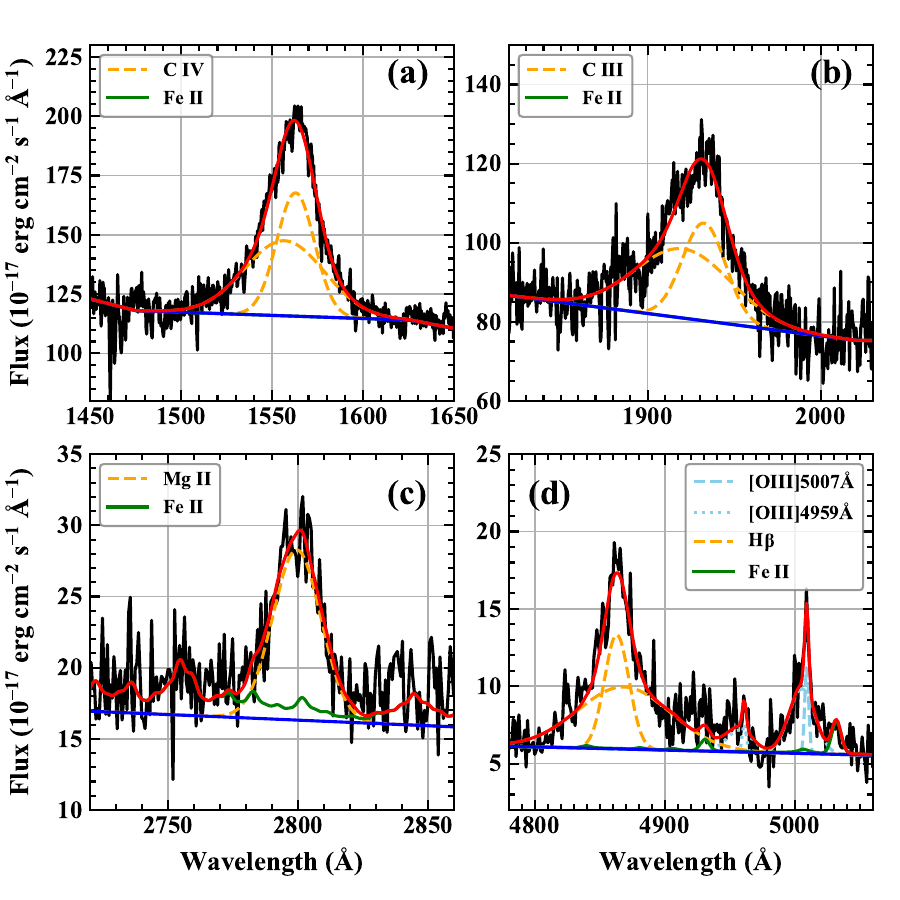}  \\
\end{tabular}
\caption{Examples of fitting some strong optical/UV emission lines with Gaussian components. Panels-a and b are for SDSS J030341.04-002321.9, and Panels-c and d are for SDSS J004250.54+010205.9. The original spectra, after Galactic reddening and redshift corrections, are shown in black. The red line is the best-fit model, the blue line is the best-fit local underlying continuum. The green is the best-fit Fe {\sc ii} lines using the template of \citet{Verner.1999}.}
\label{fig-linefit}
\end{figure}

To investigate the correlations for the emission lines and their underlying continua separately, we perform a local line profile fitting to separate the line flux from its underlying continuum. The continuum is fitted locally by a straight line, and the emission lines are fitted with multiple Gaussian components. This is performed for the C {\sc iv}, C {\sc iii}], Mg {\sc ii}, H$\beta$ and [O {\sc iii}]$\lambda$5007 lines locally, for the sources whose SDSS spectra exhibit these lines. Examples of the line-fitting results are shown in Figure~\ref{fig-linefit}. The correlations for individual lines and continua are shown in Figure~\ref{fig-linecor}, and the regression parameters are listed in Table~\ref{tab-linecor}. Some quasars in our sample also display strong absorption features in the C {\sc iv}, C {\sc iii}] lines, which can severely affect the line profile modelling. Other quasars may have incomplete line profiles because the line is located at the edge of the spectral range. We exclude these sources. This is why for some lines (C {\sc iv}, C {\sc iii}], H$\beta$) listed in Table~\ref{tab-linecor} the number of sources with a line flux is slightly smaller than for the continuum flux.

We find that the emission lines and continua are both well correlated with the X-ray luminosity. The underlying continua of all the lines show a regression slope of 0.6 -- 0.7 and intercept 6 -- 8, fully consistent with other wavelengths. But the parameters found for individual lines are indeed significantly different from the continua (see Table~\ref{tab-ouxcs-linecor}), confirming that these optical/UV lines have different correlations with the X-ray luminosity.

\begin{table}
\centering
   \caption{The LINMIX regression parameters at the central wavelengths of different emission lines in Figure~\ref{fig-linmix}. Parameter $\gamma$, $\beta$ and $\delta$ are the regression slope, intercept and dispersion, respectively.}
    \begin{tabular}{lccc}
    \hline
    Line & $\gamma$ & $\beta$ & $\delta$ \\
    \hline
     Ly$\alpha$ & 0.556 $\pm$ 0.049 & 9.70 $\pm$ 1.56 & 0.220 $\pm$ 0.018 \\\relax
     C {\sc iv}    & 0.707 $\pm$ 0.026 & 4.91 $\pm$ 0.81 & 0.216 $\pm$ 0.009 \\\relax
     C {\sc iii}]   & 0.662 $\pm$ 0.020 & 6.34 $\pm$ 0.62 & 0.231 $\pm$ 0.007 \\\relax
     Mg {\sc ii}   & 0.680 $\pm$ 0.017 & 5.67 $\pm$ 0.51 & 0.229 $\pm$ 0.006 \\\relax
     [Ne {\sc v}] & 0.651 $\pm$ 0.019 & 6.66 $\pm$ 0.58 & 0.247 $\pm$ 0.007 \\\relax
     [O {\sc ii}]   & 0.665 $\pm$ 0.023 & 6.22 $\pm$ 0.69 & 0.245 $\pm$ 0.007 \\\relax
     [Ne {\sc iii}] & 0.657 $\pm$ 0.024 & 6.46 $\pm$ 0.73 & 0.248 $\pm$ 0.008 \\\relax
     H$\gamma$ & 0.656 $\pm$ 0.032 & 6.45 $\pm$ 0.96 & 0.258 $\pm$ 0.010 \\\relax
     [O {\sc iii}] $\lambda$3728 & 0.661 $\pm$ 0.032 & 6.30 $\pm$ 0.97 & 0.255 $\pm$ 0.010 \\\relax
     He {\sc ii} & 0.685 $\pm$ 0.042 & 5.62 $\pm$ 1.26 & 0.266 $\pm$ 0.011 \\\relax
     H$\beta$ & 0.695 $\pm$ 0.046 & 5.15 $\pm$ 1.39 & 0.266 $\pm$ 0.012 \\\relax
     [O {\sc iii}] $\lambda$4959 & 0.750 $\pm$ 0.045 & 3.67 $\pm$ 1.34 & 0.246 $\pm$ 0.012 \\\relax
     [O {\sc iii}] $\lambda$5007 & 0.806 $\pm$ 0.048 & 1.85 $\pm$ 1.45 & 0.242 $\pm$ 0.012 \\
     \hline
     \end{tabular}
\label{tab-ouxcs-linecor}
\end{table}

Similar results are obtained when we correlate these lines and their underlying continua with the 2500 \AA\ luminosity (i.e. $L_{\rm 2500\textup{\AA}, phot}$). The underlying continua show consistent regression parameters with the other wavelengths, while the lines themselves show different parameters. For AGN in general, there is an anti-correlation between the continuum luminosity and the equivalent width of broad emission lines, which is known as the Baldwin effect (\citealt{Baldwin.1977}). Such an anti-correlation can be found for various optical/UV broad emission lines, although the slope can change significantly with the ionizing potential (\citealt{Zheng.1993, Dietrich.2002}). For example, \citet{Dietrich.2002} used the luminosity at 1450 \AA\ as the continuum luminosity, and then reported the slope to be $-$0.20 $\pm$ 0.03 for C {\sc iv}, $-$0.09 $\pm$ 0.02 for C {\sc iii}], $-$0.09 $\pm$ 0.01 for Mg {\sc ii} and 0.01 $\pm$ 0.03 for H$\beta$. Note that in the Baldwin effect the continuum luminosity is used as the independent variable. For our sample we treat $L_{\rm 2500\textup{\AA}, phot}$ as the independent variable and derive the correlation and slope for individual lines (see Figure~\ref{fig-baldwin}). We find the slope to be $-$0.14 $\pm$ 0.04 for C {\sc iv}, $-$0.08 $\pm$ 0.03 for C {\sc iii}], $-$0.08 $\pm$ 0.01 for Mg {\sc ii} and 0.10 $\pm$ 0.02 for H$\beta$. Therefore, with our large and clean quasar sample, we confirm the slopes of the Baldwin effect for various broad emission lines as reported by \citet{Dietrich.2002}. 
Furthermore, we find the slope to be $-$0.16 $\pm$ 0.05 for [O {\sc iii}] $\lambda$5007 line, which was not reported by \citet{Dietrich.2002} for the high-luminosity range.

\section{Discussion}
\label{sec-discussion}
\subsection{Is 2500 \AA\ Representative for the Optical/UV Continuum?}
\label{sec-2500a}
Based on the results presented in the previous sections we can address  the question proposed at the beginning of this paper: whether the conventional choice of 2500 \AA\ is a good choice to represent the optical/UV luminosity and its correlation with the X-rays, for the large quasar sample. As shown by the OUXCS-2 in Figure~\ref{fig-oxcs-2} Panel-a2 , there is a strong correlation with the X-ray luminosity at every wavelength within 1280 -- 5550\AA\ if the spectral quality is high. The correlation coefficients found at other wavelengths are consistent with those at 2500 \AA. The weak wavelength-dependence of the regression parameters shown in Figure~\ref{fig-linmix} also suggest that the slope and intercept found at 2500 \AA\ are similar to those found at the other wavelengths. The regression parameters only change significantly at the wavelengths covering strong emission lines. But after these lines are subtracted, their underlying continua still show similar correlations as seen at other wavelengths. Therefore, we conclude that results of optical/UV and X-ray inter-band correlations found by previous studies using 2500 \AA\, are indeed robust and representative for the entire optical/UV continuum, provided that the quasar sample is composed of unobscured objects with high-quality data. Our results also highlight the efficiency of the filtering steps we applied for selecting unobscured quasar samples, as described in \citet{Lusso.2020}.

\subsection{Various Measurements of $L_{\rm 2500\textup{\AA}}$}
\label{sec-l2500var}
The luminosity at the rest frame 2500 \AA\ ($L_{\rm 2500\textup{\AA}}$) can be measured in different ways: from the photometric spectral energy distribution (i.e. $L_{\rm 2500\textup{\AA}, phot}$, thus including the contribution of both the continuum and emission/absorption lines), or directly from the spectrum (i.e. $L_{\rm 2500\textup{\AA}, spec}$). Note that there are only 902 sources whose SDSS spectra cover 2500 \AA\ in the rest-frame, so that only this subsample has $L_{\rm 2500\textup{\AA}, spec}$ measurement.

\begin{figure*}
\centering
\begin{tabular}{ccc}
\includegraphics[trim=0.1in 0.2in 0.0in 0.0in, clip=1, scale=0.45]{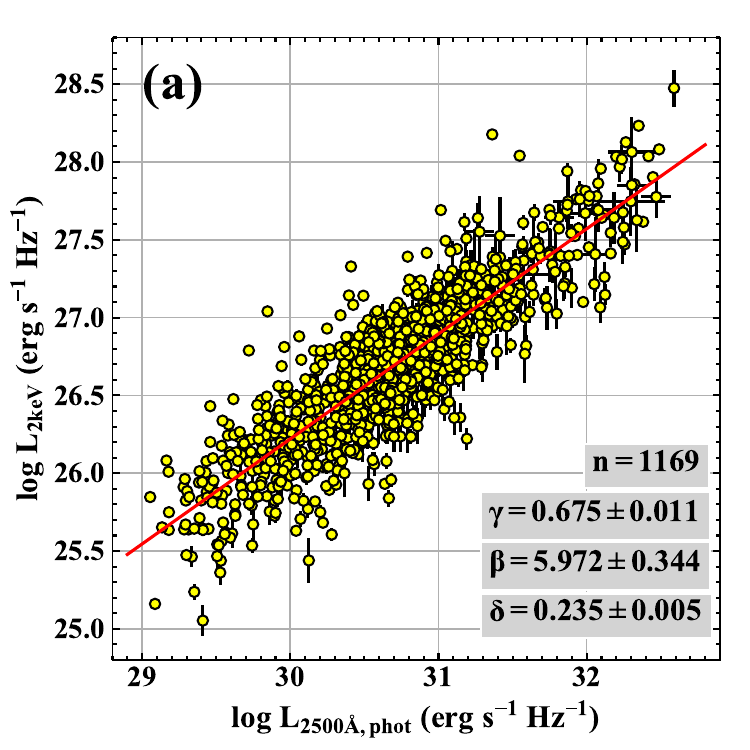}  &
\includegraphics[trim=0.2in 0.2in 0.0in 0.0in, clip=1, scale=0.45]{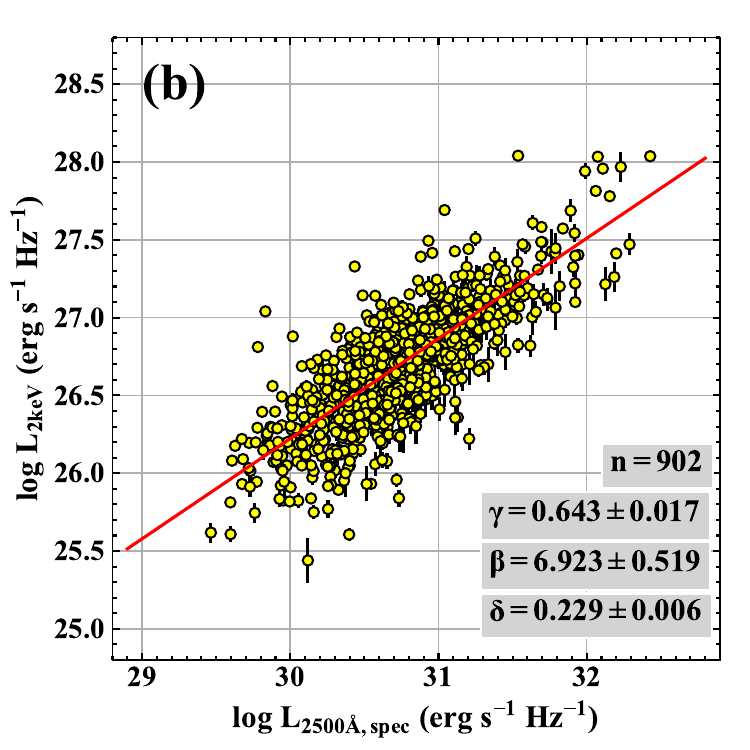}  &
\includegraphics[trim=0.2in 0.2in 0.0in 0.0in, clip=1, scale=0.45]{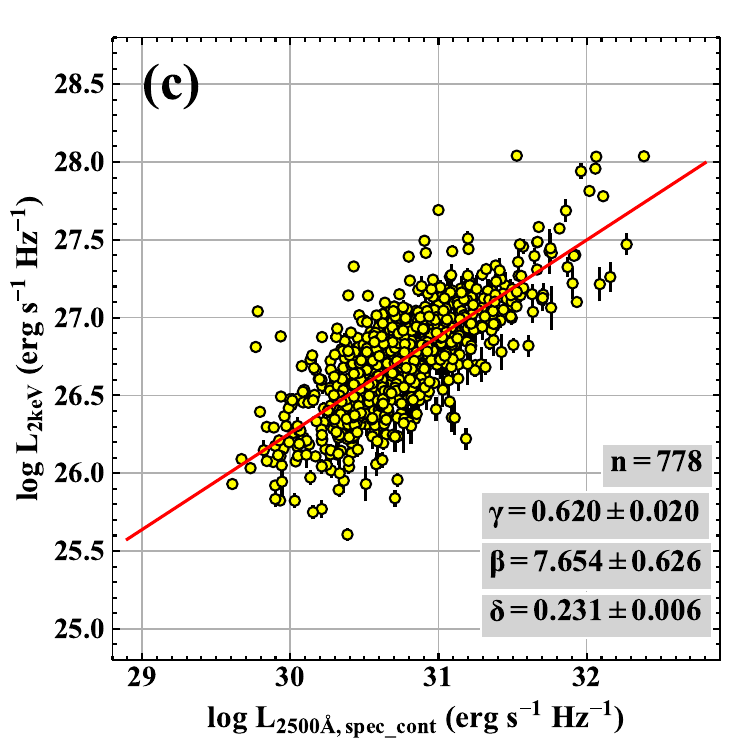}  \\
\end{tabular}
\caption{Comparison of the regression results for the $L_{\rm 2500\textup{\AA}}$ vs. $L_{\rm 2 keV}$ relation for different measurements of $L_{\rm 2500\textup{\AA}}$, including the photometric $L_{\rm 2500\textup{\AA}, phot}$ in Panel-a, spectroscopic $L_{\rm 2500\textup{\AA}, spec}$ in Panel-b and the spectral continuum $L_{\rm 2500\textup{\AA}, cont}$ in Panel-c. $n$ is the number of sources in each subsample.}
\label{fig-checkcor}
\end{figure*}

For our sample, the systematic difference between $L_{\rm 2500\textup{\AA}, phot}$ and $L_{\rm 2500\textup{\AA}, spec}$ is only 6.6 per cent (see Figure~\ref{fig-2500cross}), so their regression results are expected to be similar. Indeed, Figures~\ref{fig-checkcor}a and \ref{fig-checkcor}b show the regression results for the two measurements of $L_{\rm 2500\textup{\AA}}$. No significant difference is found between the regression parameters, including a similar level of intrinsic dispersion.

\begin{figure}
\centering
\begin{tabular}{c}
\includegraphics[trim=0.1in 0.2in 0.0in 0.0in, clip=1, scale=0.55]{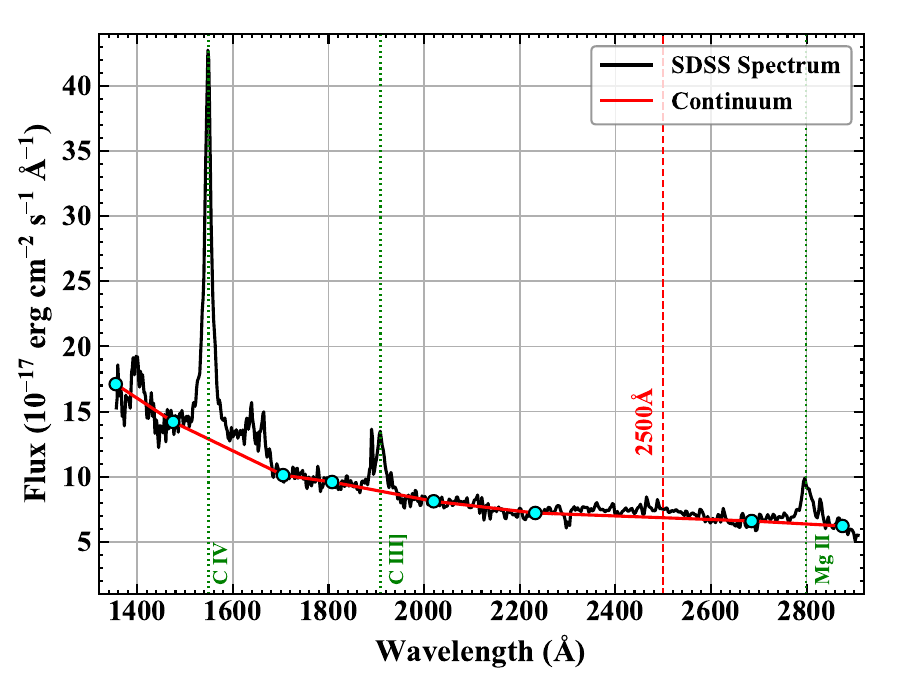}  \\
\end{tabular}
\caption{The SDSS spectrum of the quasar SDSS J013723.81-082133.9 (de-reddened and de-redshifted). Cyan circles indicate the line-free positions used to define the underlying continuum (red solid line).}
\label{fig-conti}
\end{figure}

We note that the flux at 2500 \AA\ contains both the continuum and the blend of Fe {\sc ii} UV emission lines (\citealt{Vestergaard.2001}), so it is also useful to examine whether the UV Fe {\sc ii} emission changes the regression results at 2500 \AA\ significantly. For this purpose, we use a series of line-free wavelengths to determine the UV continuum, as shown in Figure~\ref{fig-conti}. We then visually inspect each spectrum to ensure the quality. Then $L_{\rm 2500\textup{\AA}}$ is measured from the continuum (i.e. $L_{\rm 2500\textup{\AA}, cont}$). In most cases we use the flux at 2230 \AA\ as one continual point, so it results in a reduction in the sample size. Some sources are also excluded due to the poor quality on the blue end of the spectrum. Thus the final subsample with $L_{\rm 2500\textup{\AA}, cont}$ measurement has only 778 sources. We find that $L_{\rm 2500\textup{\AA}, spec}$ is statistically larger than $L_{\rm 2500\textup{\AA}, cont}$ by only 6.4 $\pm$ 5.4 per cent. Figures~\ref{fig-checkcor}c shows the regression result for $L_{\rm 2500\textup{\AA}, cont}$, and the results are still consistent with the other two measurements of $L_{\rm 2500\textup{\AA}}$, including a similar intrinsic dispersion of 0.231 $\pm$ 0.006 dex.

Therefore we conclude that the regression results for the $L_{\rm 2500\textup{\AA}}$ vs. $L_{\rm 2 keV}$ relation are not affected by different measurements of $L_{\rm 2500\textup{\AA}}$ in this study.

\subsection{The $L_{\rm 2 keV}$ -- $L_{\rm 2500\textup{\AA}}$ -- $\nu_{\rm fwhm}$ Plane}
\label{sec-vfwhm}
\begin{figure}
\centering
\begin{tabular}{c}
\includegraphics[trim=0.4in 0.3in 0.0in 0.0in, clip=1, scale=0.58]{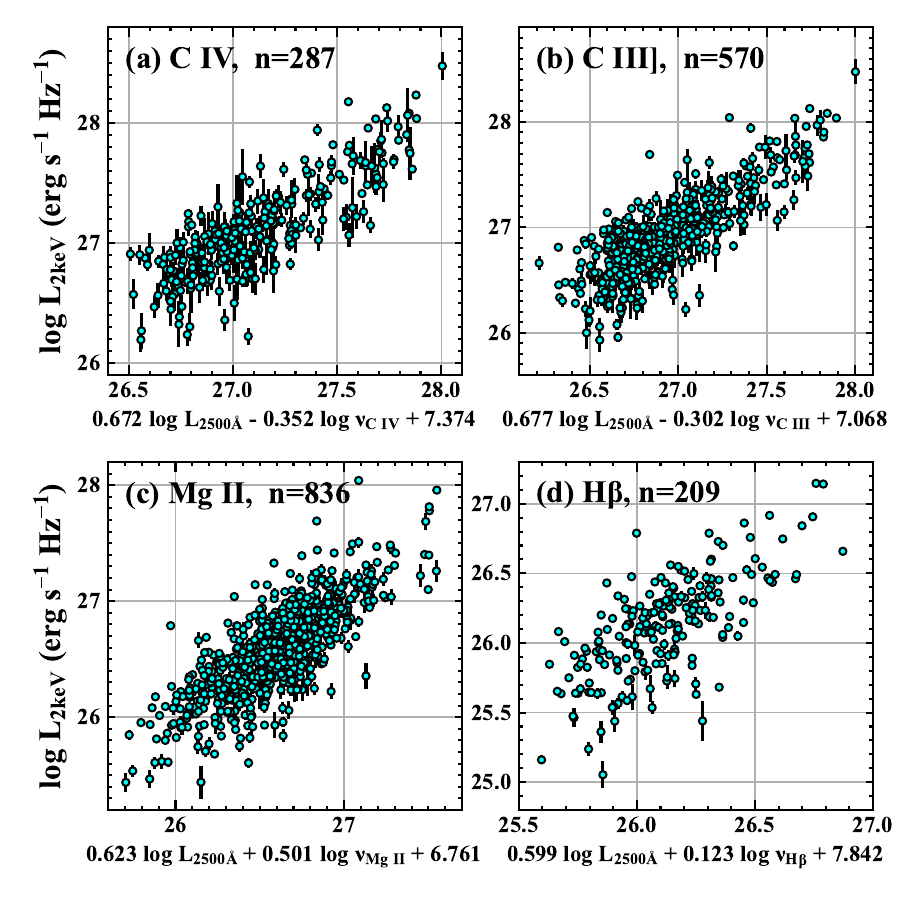}  \\
\end{tabular}
\caption{The $L_{\rm 2 keV}$ -- $L_{\rm 2500\textup{\AA}}$ -- $\nu_{\rm fwhm}$ plane seen edge-on for different emission lines.  $n$ is the number of sources in each subsample.}
\label{fig-2dcor}
\end{figure}

\citet{Lusso.2017} adopted a simplified disc-corona model and assumed that the radius of the broad line region (BLR) is proportional to the square-root of the disc luminosity, and then proposed the theoretical relation: $L_{\rm 2keV}\propto L_{\rm optuv}^{4/7}\nu_{\rm fwhm}^{4/7}$, where $\nu_{\rm fwhm}$ is the virial velocity of the BLR. This relation was confirmed by their statistical analysis of a sample of 545 optically selected quasars. In this work, the Mg {\sc ii} line was used to measure $\nu_{\rm fwhm}$. Since line-profile fitting has been conducted for all the sources (see Section~\ref{sec-wdep}), we can perform 2-dimensional (2D) regression in the $L_{\rm 2 keV}$ -- $L_{\rm 2500\textup{\AA}}$ -- $\nu_{\rm fwhm}$ plane to test the above theoretical relation. This 2D regression is defined as:
\begin{equation}
\label{eq-vfwhm}
{\rm log}(L_{\rm 2~keV}) = \gamma_{1} {\rm log}(L_{\rm 2500~\textup{\AA}}) + \gamma_{2} {\rm log}(\nu_{\rm fwhm}) + \beta_{1}
\end{equation}
where $\gamma_{1}$, $\gamma_{2}$ and $\beta_{1}$ are the free parameters to be fitted. Thus we have $\gamma_{1}$ and $\gamma_{2}$ equal to 4/7 for the theoretical relation. We also define $\delta_{1}$ as the intrinsic dispersion of this plane, to distinguish it from the intrinsic dispersion ($\delta$) of the $L_{\rm 2 keV}$ -- $L_{\rm 2500\textup{\AA}}$ relation. We conducted the 2D regression using the $\nu_{\rm fwhm}$ of C {\sc iv}, C {\sc iii}], Mg {\sc ii} and H$\beta$, respectively. The coefficients are listed in Table~\ref{tab-2dcor}. The planes seen edge-on are shown in Figure~\ref{fig-2dcor}.

The subsample of Mg {\sc ii} has 836 quasars, each has measurements of $L_{\rm 2 keV}$, $L_{\rm 2500\textup{\AA}, phot}$ and Mg {\sc ii} line width. We find $\gamma_{2}=0.623\pm0.017$ and $\gamma_{2}=0.501\pm0.046$, which are statistically consistent with both the theoretical values and the results reported by \citet{Lusso.2017} for the same line but a different, larger sample. We also find that, for the same subsample, the dispersion of the $L_{\rm 2 keV}$ -- $L_{\rm 2500\textup{\AA}}$ relation decreases by $0.012\pm0.003$ dex when the $\nu_{\rm fwhm}$ of Mg {\sc ii} is included in the regression (see Table~\ref{tab-2dcor}).

For the H$\beta$ line, the subsample of 209 quasars displays a relatively small dynamical range in luminosity (most of the data lies in the range 27.5 -- 28.5, $\sim$1 dex, see Figure~\ref{fig-linecor}), and the intrinsic dispersion of the relation appears to dominate the distribution (see Figure~\ref{fig-2dcor}d). We find a similar index of $\gamma_{1}=0.627\pm0.045$ for $L_{\rm 2500\textup{\AA}, phot}$. But $\gamma_{2}$ is found to be $0.165\pm0.162$. However the large uncertainty makes it difficult to draw any firm conclusion about the index of $\nu_{\rm fwhm}$.

The results for C {\sc iv} and C {\sc iii}], however, are significantly different. We find $\gamma_{2}$ is $-0.352\pm0.067$ for C {\sc iv} and $-0.302\pm0.049$ for C {\sc iii}], both of which are negative, and opposite to the indices found for Mg {\sc ii}, H$\beta$ and the theoretical value. The negative $\gamma_{2}$ indicates that larger line-widths of C {\sc iv} and C {\sc iii}] correspond to weaker X-ray luminosity relative to UV. Considering the fact that the line-widths of C {\sc iv} and C {\sc iii}] do not only include components of virial velocity, but also a significant outflow velocity of the BLR (\citealt{Gaskell.1982, Wilkes.1984, Leighly.2004}), the negative $\gamma_{2}$ is actually consistent with the anti-correlation found between the optical-to-X-ray spectral index ($\alpha_{\rm ox}$: \citealt{Tananbaum.1979, Lusso.2010}) and the blueshift velocity of C {\sc iv} (\citealt{Timlin.2020, Lusso.2021}). It is well known that AGN properties exhibit different sets of correlations. The primary set is the so-called `Eigenvector-1' (\citealt{Boroson.1992}), which is found to be driven by the Eddington ratio (e.g. \citealt{Marziani.2003, Grupe.2004, Jin.2012c}). The blueshift of C {\sc iv} also belongs to Eigenvector-1, and a higher Eddington ratio corresponds to a larger blueshift velocity of this line (\citealt{Sulentic.2007}). In addition, it is found that a higher Eddington ratio corresponds to a lower ratio of X-ray emission relative to UV (e.g. \citealt{Vasudevan.2007, Lusso.2010, Grupe.2010, Jin.2012c}). Therefore, these previously known correlations with the Eddington ratio can naturally result in a  negative $\gamma_{2}$ of C {\sc iv} and C {\sc iii}].

\begin{figure*}
\centering
\begin{tabular}{cc}
\includegraphics[trim=0.2in 0.2in 0.0in 0.0in, clip=1, scale=0.53]{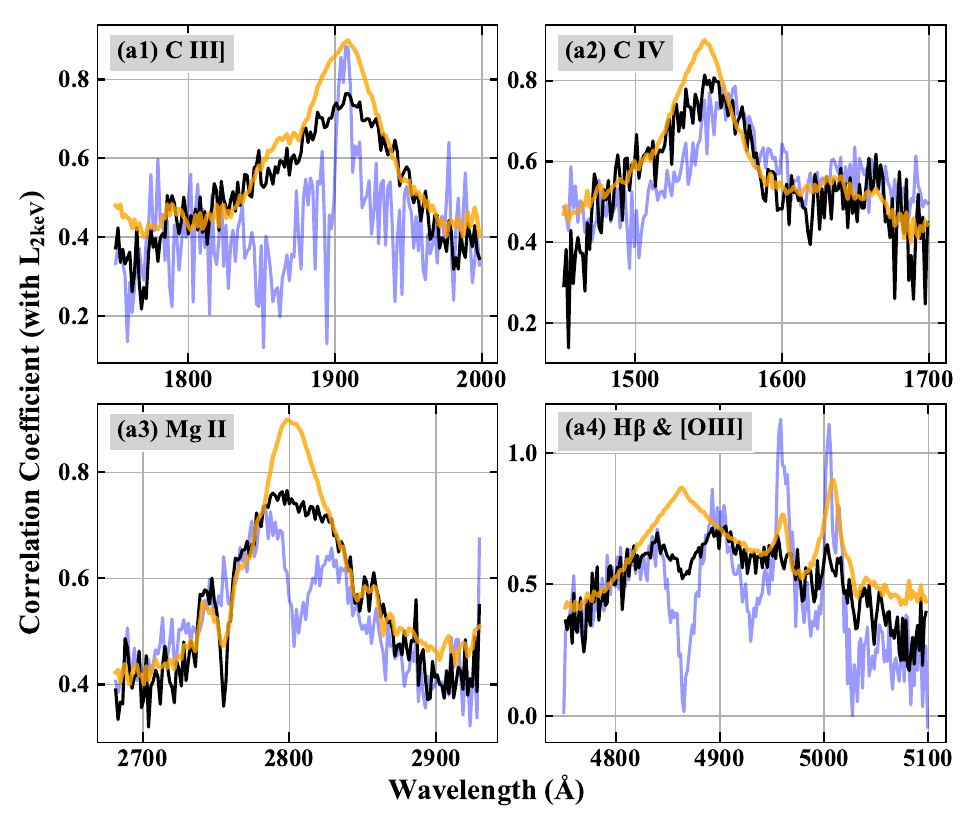} &
\includegraphics[trim=0.1in 0.2in 0.0in 0.0in, clip=1, scale=0.53]{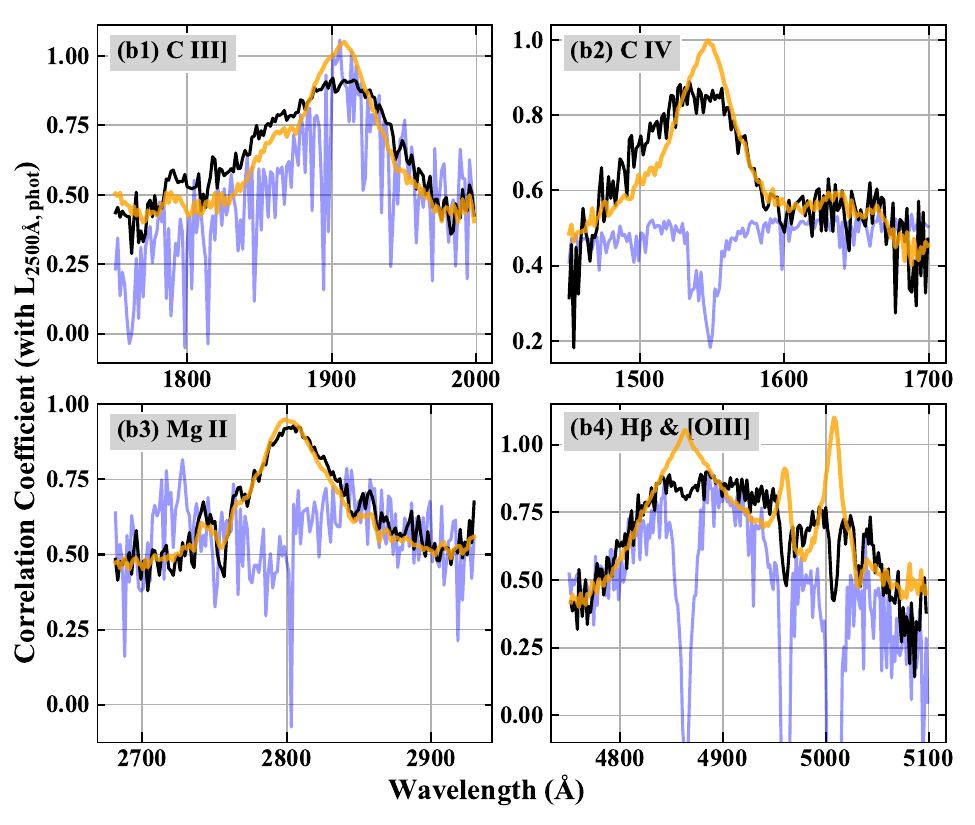} \\
\end{tabular}
\caption{Direct correlation spectra for different emission lines after the subtraction of underlying continuum and Fe {\sc ii} lines (plotted in black in each panel). The composite line profile (rescaled and renormalized, plotted in orange) and the correlation spectra before continual subtraction (rescaled and renormalized, plotted in light blue) are shown for the line-profile comparison only. The four panels on the left  (a1, a2, a3, a4) are for correlations with $L_{\rm 2 keV}$, while the four panels on the right  (b1, b2, b3, b4) are for correlations with $L_{\rm 2500\textup{\AA},phot}$.}
\label{fig-ouxcs-line}
\end{figure*}

\begin{table*}
\centering
   \caption{Comparison of the regression results for the $L_{\rm 2 keV}$ -- $L_{\rm 2500\textup{\AA}}$ relation and the $L_{\rm 2 keV}$ -- $L_{\rm 2500\textup{\AA}}$ -- $\nu_{\rm fwhm}$ plane for different emission lines. The parameters are defined in Equations~\ref{eq-luvx} and \ref{eq-vfwhm}. $n$ is the number of sources in each subsample. $\delta$ and $\delta_{1}$ are the intrinsic dispersion for the two types of regression.}
    \begin{tabular}{@{}lcccccccc@{}}
     \hline
    $\nu_{\rm fwhm}$ & $n$ & $\gamma$ & $\beta$ & $\delta$ & $\gamma_{1}$ & $\gamma_{2}$ & $\beta_{1}$ & $\delta_{1}$ \\
     \hline
     C {\sc iv} & 287 & 0.658 $\pm$ 0.027 & 6.52 $\pm$ 0.80 & 0.209 $\pm$ 0.006 & 0.672 $\pm$ 0.024 & -0.352 $\pm$ 0.067 & 7.37 $\pm$ 0.80 & 0.198 $\pm$ 0.006 \\
     C {\sc iii}] & 570 & 0.654 $\pm$ 0.020 & 6.64 $\pm$ 0.62 & 0.213 $\pm$ 0.003 & 0.677 $\pm$ 0.019 & -0.302 $\pm$ 0.049 & 7.07 $\pm$ 0.67 & 0.206 $\pm$ 0.003 \\
     Mg {\sc ii} & 836 & 0.646 $\pm$ 0.018 & 6.84 $\pm$ 0.55 & 0.237 $\pm$ 0.002 & 0.623 $\pm$ 0.017 & 0.501 $\pm$ 0.046 & 6.76 $\pm$ 0.52 & 0.225 $\pm$ 0.002 \\
     H$\beta$ & 209 & 0.643 $\pm$ 0.043 & 6.94 $\pm$ 1.23 & 0.252 $\pm$ 0.002 & 0.627 $\pm$ 0.045 & 0.165 $\pm$ 0.162 & 7.06 $\pm$ 1.30 & 0.252 $\pm$ 0.002  \\
     \hline
     \end{tabular}
\label{tab-2dcor}
\end{table*}

\subsection{Comparing the Direct-Correlation Method with the Line-fitting Method}
\label{sec-compare-method}
The OUXCS and OUSCS produced by the direct-correlation method shows that, compared to the underlying continuum, the primary emission lines have different correlations with the 2 keV and 2500 \AA\ luminosities. This is consistent with the results found by fitting the lines, separating the line flux from the continuum, and then comparing their individual correlations. However, these two methods have fundamentally different characteristics.

The line-fitting method can separate the emission lines from the continuum, and test their correlations separately with some other parameters (such as 2 keV luminosity and 2500\AA\ luminosity), but this method may also introduce the following issues.

Firstly, the uncertainty caused by the line profile fitting. There are various complexities in the line profile fitting. For example, the broad and narrow components are blended; the broad H$\beta$, Fe {\sc ii} and [O {\sc iii}] are also blended; narrow absorption lines can be present with different velocity-shifts in C {\sc iv} and C {\sc iii}], although these absorption features are not visible in the mean spectrum in Figure~\ref{fig-wdist}d. These can all lead to a measurement uncertainty of the line flux. However, the results for individual lines in Section~\ref{sec-wdep} are not affected by the line decomposition because the total line flux is used.

Secondly, the relative error of the line flux. The LINMIX algorithm is sensitive to the relative errors of the data points. Once the lines are separated from the continuum, the relative errors of the line fluxes increase, especially when the lines are weak. This can lead to different regression parameters. 

Thirdly, different correlations at different velocity shifts of one line. For example, C {\sc iv} has a significant blue wing, but OUXCS does not show a similarly strong correlation in the blue side of the line. This means that the correlation of the entire line flux will be the mean value of the correlations for components present at all velocity shifts.

In comparison, the direct-correlation method examines the correlation with a specific parameter (such as 2 keV luminosity or 2500\AA\ luminosity) at each wavelength. At the wavelength of an emission line, the correlation given is the overall result of the emission line, the underlying continuum and nearby blended lines (such as the Fe {\sc ii} lines). If the correlation of an emission line shows a similar trend to that of the continuum, but with a statistically stronger correlation, it appears as an `emission line' in the correlation spectrum as well, just like C {\sc iv}, C {\sc iii}], Mg {\sc ii}, H$\beta$ in OUXCS. Conversely, if the correlation of the line is different from or weaker than that of the continuum, it appears as an `absorption line' in the correlation spectrum, just like [O {\sc iii}], C {\sc iv}, Mg {\sc ii} and the narrow H$\beta$ in OUSCS. Therefore, the direct-correlation method can be used to study the difference in correlation between emission lines and the continuum without detailed spectral fitting.

The advantage of this method can be reflected by the correlation parameters seen at the optical Fe {\sc ii} complex in the OUXCS (Figures~\ref{fig-oxcs-2} Panel-a2 and \ref{fig-linmix}). Usually, it is very difficult to explore the correlation of Fe {\sc ii} lines with other parameters unless they are very strong (e.g. in some super-Eddington NLS1s, \citealt{Du.2019}) and not blended with nearby lines (e.g. [O {\sc iii}] $\lambda\lambda$4959/5007). With the direct-correlation method, we can reveal the correlation at the Fe {\sc ii} complex without line fitting or de-blending. However, the disadvantage is that if different lines in a certain waveband are mixed together, the result of direct correlation will contain the correlations of different lines, making it difficult to distinguish which line has a greater contribution.

We can also apply the direct-correlation method to each emission line. Based on the line fitting described in Section~\ref{sec-wdep}, we first remove the best-fit local continuum and Fe {\sc ii} lines, and then directly correlate the spectral lines with the 2 keV luminosity and 2500\AA\ luminosity, respectively. Figure~\ref{fig-ouxcs-line} shows the direct correlation results for different emission lines (plotted in black in each panel). Firstly, by comparing with the composite line profile (plotted in orange), we find that the correlation between the narrow line component and 2 keV is weaker than that of the broad line component. Secondly, the direct correlation results of the line profiles are very different from the line profiles observed in OUXCS and OUSCS (plotted in light blue), which indicates that the shapes of OUXCS and OUSCS are the overall results of different spectral components. For example, when we look at Panels a1, a2, a3 and a4 separately, we find that the blue wing of C {\sc iv} and C {\sc iii}], as well as the narrow component of Mg {\sc ii} and H$\beta$, are relatively weaker in OUXCS than those observed in the direct correlation line profiles, indicating that the correlations of these line components are more different from the continuum.

Therefore, we can see that the direct correlation method is simple and straightforward without introducing the complexities of line fitting, and it also provides new clues for understanding the profile and origin of different lines which we discuss in more detail in the next section.

\subsection{Origins of Different Line Species and Line Components}
\label{sec-linecor}
The direct-correlation line profiles in Figure~\ref{fig-ouxcs-line} show the correlation of emission line at different velocities. In comparison, the lines observed in OUXCS and OUSCS show if and how significant are their correlations (with 2 keV luminosity or 2500 \AA\ luminosity) different from those of the continuum. We summarize these correlation properties and, whenever possible, explain them in terms of the AGN unified model (\citealt{Antonucci.1993}) and the shape of the spectral energy distribution (SED) of the ionizing photons , thereby inferring the origin of different lines and line components.

(1) By comparing the direct-correlation line profiles of 2 keV with 2500 \AA\ in Figure~\ref{fig-ouxcs-line}, we find that the correlations of C {\sc iv}, C {\sc iii}], Mg {\sc ii}, H$\beta$ lines with the 2500 \AA\ luminosity are systematically stronger than those with 2 keV luminosity.

These results can be understood from two aspects. Firstly, the ionization potentials of C {\sc iv}, C {\sc iii}], [O {\sc iii}] and Mg {\sc ii} are 47.9 eV, 24.4 eV, 35.1 eV and 7.6 eV, respectively. These energies correspond to wavelengths of 259.5\AA, 509.5\AA, 354.2\AA\ and 1635.7\AA. Therefore, these lines should be correlated with the far-UV and X-ray emission. The fact that they are all well correlated with the 2500 \AA\ luminosity suggests that the good optical/UV self-correlation seen in OUSCS-2 (Figure~\ref{fig-oxcs-2} Panel-b2) should extend into the far-UV regime of at least a few hundreds of angstrom, which is likely also dominated by the accretion disc emission. Secondly, X-rays are easily absorbed and only account for a low proportion of the broadband spectral energy distribution (SED). For example, based on the classic flat equatorial distribution of BLR, X-rays may be significantly absorbed by the disc wind or the disc itself before reaching the BLR. Therefore, it is reasonable to expect that these BLR lines should correlate better with the UV continuum than X-ray.

(2) By comparing the direct-correlation line profile (black) and composite line profile (orange) in each panel of Figure~\ref{fig-ouxcs-line}, we find that the correlation between the narrow component of the broad emission lines and 2 keV or 2500 \AA\ is generally worse than that of the broad component. However, the direct-correlation line profile of Mg {\sc ii} for 2500 \AA\ are almost the same as the composite line profile, indicating that the narrow component of Mg {\sc ii} also have a similarly good correlation with 2500 \AA\ luminosity.

These results can be understood as the broad component being mainly produced by the ionization of disc photons, while the narrow component may also include contaminations from the host galaxy star light or being affected by other factors (such as covering factor), so the correlation of the narrow component with 2 keV and 2500 \AA\ is worse. This is also consistent with the results of \citet{Jin.2012b}.

(3) By comparing the direction-correlation line profiles (black) in Panels a1 to a4 of Figure~\ref{fig-ouxcs-line} with OUXCS (light blue), we find that the correlation on the blue side of C {\sc iv} is somewhat different from the continuum, and the central narrow components of Mg {\sc ii} and H$\beta$ show similar behaviours. We discuss these results together with the next point below.

(4) By comparing the direction-correlation line profiles (black) and OUXCS (light blue) in Panels b1 to b4 of Figure~\ref{fig-ouxcs-line}, we find that although strong correlations with 2500 \AA\ are observed across the entire wavelength range, the lines and their continua actually behave differently, but only in the cases of high-ionization lines (i.e. C {\sc iv}, C {\sc iii}]) and narrow component of low-ionization lines (H$\beta$, Mg {\sc ii}). This is also confirmed by the linear regression parameters given in Table~\ref{tab-linecor}.

The OUSCS has shown that the optical/UV continuum is highly self-correlated. Then these strong correlations should be transmitted through the ionizing source to different emission lines. However, point-3 and 4 above indicate that the forms of correlation between different lines (or different line components) and their ionizing sources are not the same. This implies that there is more complexity in the connection between these emission line regions and their respective ionizing sources, such as the difference of covering factor. For example, high-ionization lines such as C {\sc iv} are considered to be located closer to the ionizing source (e.g. \citealt{Grier.2019}), and so are more sensitive to the change of inner disc structure and wind (e.g. \citealt{Jin.2022}).

(5) Comparing to the results above, [O {\sc iii}] is a clear exception. OUXCS shows that its correlation with 2 keV is significantly better than that of the continuum and broad lines. This is consistent with previous studies of [O {\sc iii}] vs. X-ray correlation for various AGN samples (e.g. \citealt{Heckman.2005, Panessa.2006, Lamastra.2009, Jin.2012a, Jin.2012b, Ueda.2015}). Meanwhile, OUSCS shows that the correlation between [O {\sc iii}] and 2500 \AA\ is significantly worse. However, in Figure~\ref{fig-ouxcs-line} and Table~\ref{tab-linecor}, the correlation between [O {\sc iii}] and 2 keV is not significantly better. This is likely due to the additional uncertainties introduced by removing the best-fit Fe {\sc ii} lines and the underlying continuum. Thus it also demonstrates the advantage of OUXCS, which is that it does not require spectral fitting.

Compared with the narrow component of broad lines, the distribution of [O {\sc iii}]-emitting gas may be more spherical, i.e. having a significantly larger covering factor to the high-energy emission from the corona. As a result [O {\sc iii}] lines indeed show stronger correlations with the X-rays than C {\sc iv}, but weaker correlations with the optical/UV continua.

(6) The optical Fe {\sc ii} complex on both sides of H$\beta$ seems to show different correlation with e.g. the broad H$\beta$. It is known that the intensity of optical Fe {\sc ii} is strongly correlated with the Eddington ratio (i.e. Eigenvector-1, \citealt{Boroson.1992}), so it does not correlate with the optical luminosity alone, but also with the black hole mass. Furthermore, the origin of optical Fe {\sc ii} is still not clear (\citealt{Wilkes.1987, Hu.2015, Lawrence.1997, Du.2019}). It may not simply be the photoionization, but may also involve collisional excitation of gas turbulence (e.g. \citealt{Baldwin.2004}). Thus it is not surprising if the Fe {\sc ii} lines show a weaker/different correlation with the continuum luminosity.

The above explanations about different line correlations are very preliminary, but still qualitatively consistent with the classic picture of AGN unified model and BLR. However, each line requires more detailed analysis in order to fully understand their different correlations and origins, but this is beyond the scope of this work.

\section{Conclusions}
\label{sec-conclusion}
In this study we assembled a large unobscured quasar sample covering a redshift range of 0.13 -- 4.51, and applied the direct-correlation-spectrum method of \citet{Jin.2012b} to examine the optical/UV and X-ray luminosity correlations at different optical/UV wavelengths. The main results are summarized below:

\begin{itemize}
\item We presented two types of correlation spectra (OUXCS and OUSCS), and studied the wavelength dependences of the regression parameters (slope, intercept and dispersion).

\item We find that the correlations with the 2 keV and 2500 \AA\ luminosities are very significant right across the waveband of 1280 -- 5550 \AA\ when the quality of the spectral data is high. The correlations with the UV continuum are stronger than with the optical.

\item We find that the regression slope of the correlation with the 2 keV luminosity is 0.6 -- 0.7 and the intercept is 6 -- 8 for the entire optical/UV continuum, which is fully consistent with those found at 2500 \AA. Therefore, we confirm that 2500 \AA\ is indeed a good representative choice to study the optical/UV and X-ray inter-band correlations of quasars.

\item Our regression results are robust for different measurements of $L_{\rm 2500\textup{\AA}}$, such as the photometric luminosity, spectroscopic luminosity and spectral continuum luminosity.

\item The primary optical/UV emission lines (C {\sc iv},  C {\sc iii}], Mg {\sc ii}, H$\beta$, [O{\sc iii}] $\lambda\lambda$4959/5007) all show good correlations with the 2 keV and 2500 \AA\ luminosities, and the correlations with the 2500 \AA\ luminosity is systematically better. Baldwin effect have also been verified for these lines. These line correlations have different forms, and are different from their underlying continua, suggesting various complexities in the line-generation process. We provide preliminary explanations in the standard disc-wind scenario for these results.

\item We also performed 2D regression of the $L_{\rm 2 keV}$ -- $L_{\rm 2500\textup{\AA}}$ -- $\nu_{\rm fwhm}$ plane for C {\sc iv},  C {\sc iii}], Mg {\sc ii} and H$\beta$ lines, separately. The inclusion of $\nu_{\rm fwhm}$ in the regression slightly reduces the dispersion of the $L_{\rm 2 keV}$ -- $L_{\rm 2500\textup{\AA}}$ relation. We find that for Mg {\sc ii} and H$\beta$ the index $\gamma_{2}$ of $\nu_{\rm fwhm}$ is positive and statistically consistent with the theoretical value. But the indices for C {\sc iv} and C {\sc iii}] are negative, which are consistent with previously known correlations with the Eddington ratio.

\item Compared with the line-fitting method, we demonstrated the advantages of the direct-correlation method, such as its model-independence and capability to reveal correlations with different velocity components present in an emission line's profile. However, spectral fitting is still required if we want to separate the line correlation from the underlying continuum, and try to understand which spectral feature contributes the most to the correlation, especially in the wavebands where multiple spectral features are mixed together.
\end{itemize}

The difference of correlation between the observed UV continuum and the standard disc model supports with the recent work of \citet{Mitchell.2023}. Since our sample contains more quasars with higher redshifts and larger black hole masses, we will present more detailed SED analysis on this sample to further investigate the issue of UV continuum. This will also include the presence of soft excesses in some cases which may influence the luminosities in both UV and X-ray bands.

\section*{Acknowledgements}
We thank the anonymous referee for thorough reading and providing valuable comments and suggestions. CJ acknowledges the National Natural Science Foundation of China through grant 11873054, and the support by the Strategic Pioneer Program on Space Science, Chinese Academy of Sciences through grant XDA15052100. EL acknowledges the support of grant ID: 45780 Fondazione Cassa di Risparmio Firenze. CD acknowledges the Science and Technology Facilities Council (STFC) through grant ST/T000244/1 for support. MJ acknowledges support from a Leverhulme Emeritus Fellowship, EM-2021-064.

This work is based on observations conducted by \xmm, an ESA science mission with instruments and contributions directly funded by ESA Member States and the USA (NASA). This work is also based on observations made by the {\it Chandra} X-ray Observatory, as well as data obtained from the {\it Chandra} Data Archive. Funding for the SDSS and SDSS-II was provided by the Alfred P. Sloan Foundation, the Participating Institutions, the National Science Foundation, the U.S. Department of Energy, the National Aeronautics and Space Administration, the Japanese Monbukagakusho, the Max Planck Society, and the Higher Education Funding Council for England. The SDSS was managed by the Astrophysical Research Consortium for the Participating Institutions.


\section*{Data Availability}
The data underlying this article are publicly available from the High Energy Astrophysics Science Archive Research Center (HEASARC) at \url{https://heasarc.gsfc.nasa.gov}, the \xmm\ Science Archive (XSA) at \url{https://www.cosmos.esa.int/web/xmm-newton/xsa}, the {\it Chandra} Data Archive at \url{https://cda.harvard.edu/chaser}, the Sloan Digital Sky Survey (SDSS) at \url{http://classic.sdss.org/dr7/products/spectra}, and the Strasbourg astronomical Data Center at  \url{https://cds.u-strasbg.fr}.







\appendix

\section{Supplementary Figures}
\begin{figure*}
\centering
\begin{tabular}{cc}
\includegraphics[trim=0.2in 0.2in 0.0in 0.1in, clip=1, scale=0.56]{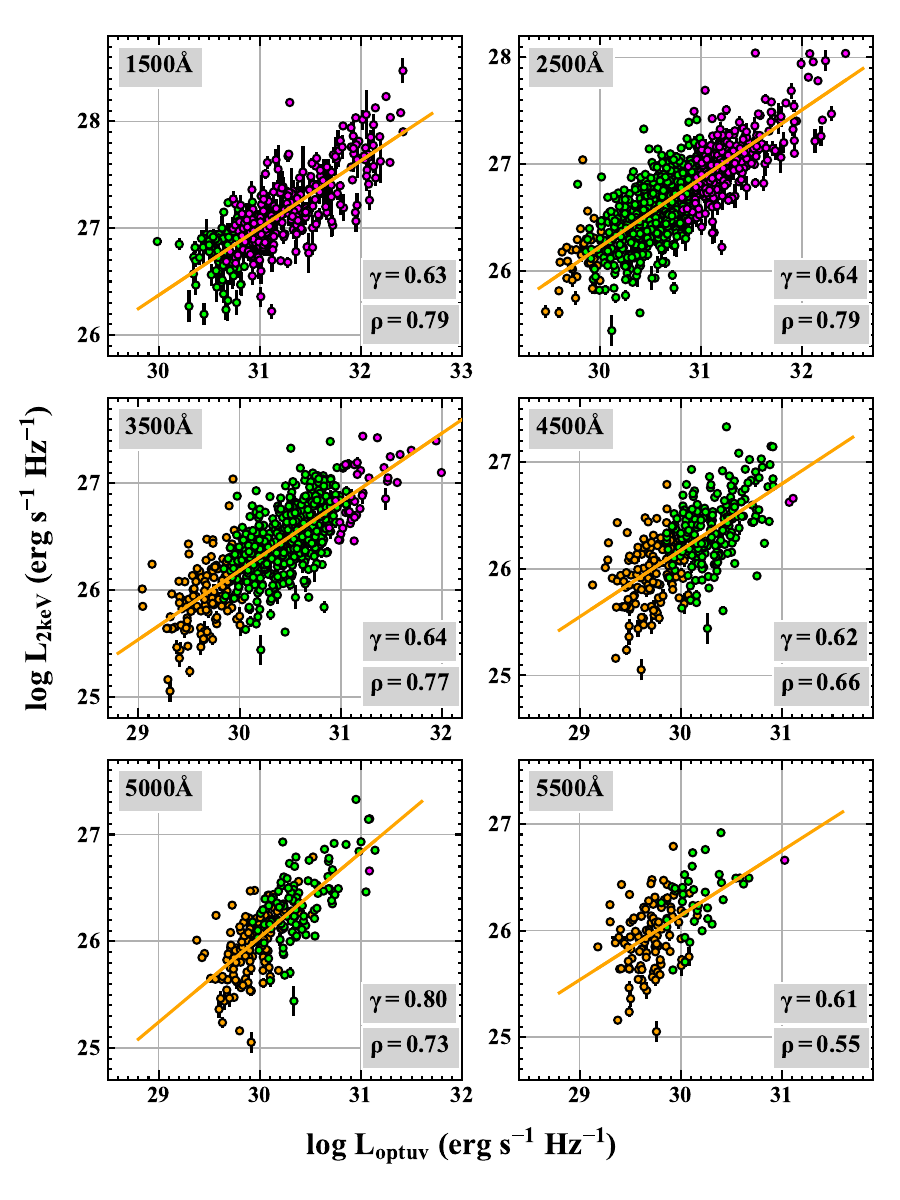} &
\includegraphics[trim=0.2in 0.2in 0.0in 0.1in, clip=1, scale=0.56]{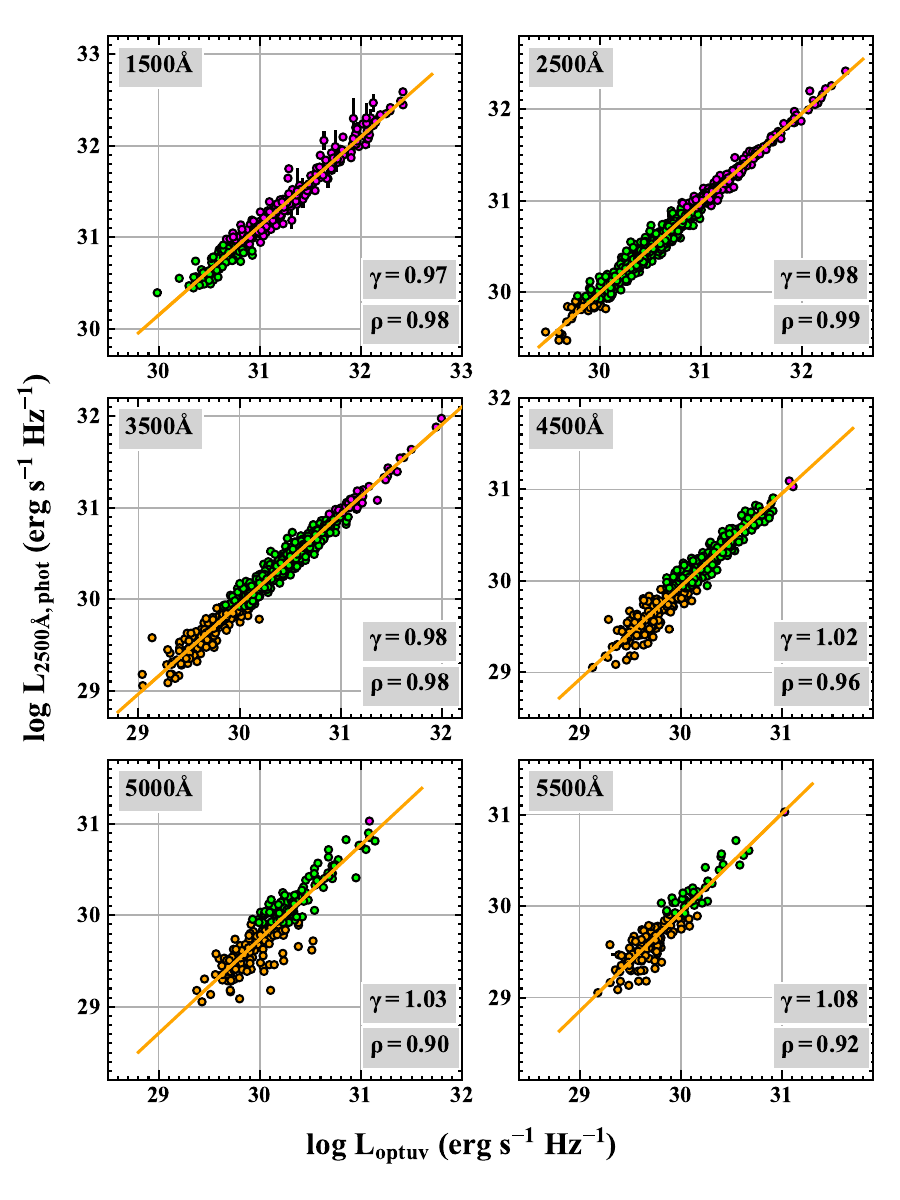} \\
\end{tabular}
\caption{Correlations for the monochromatic luminosities at different wavelengths. The left two columns are the results for correlation with the X-ray luminosity at 2 keV, and the right two columns are for the luminosity at 2500 \AA\ from photometry. In every panel, the orange line indicates the regression line found by LINMIX. $\gamma$ and $\rho$ labeled in every panel are the correlation's slope and the Pearson's correlation coefficient. The samples are divided into subsamples for different UV luminosity ranges, including $L_{\rm 2500\textup{\AA},phot}\le10^{45}$ erg s$^{-1}$ (orange),  $10^{45-46}$ erg s$^{-1}$ (green) and $>10^{46}$ erg s$^{-1}$ (magenta).}
\label{fig-contcor}
\end{figure*}

\begin{figure*}
\centering
\begin{tabular}{cc}
\includegraphics[trim=0.2in 0.2in 0.0in 0.1in, clip=1, scale=0.65]{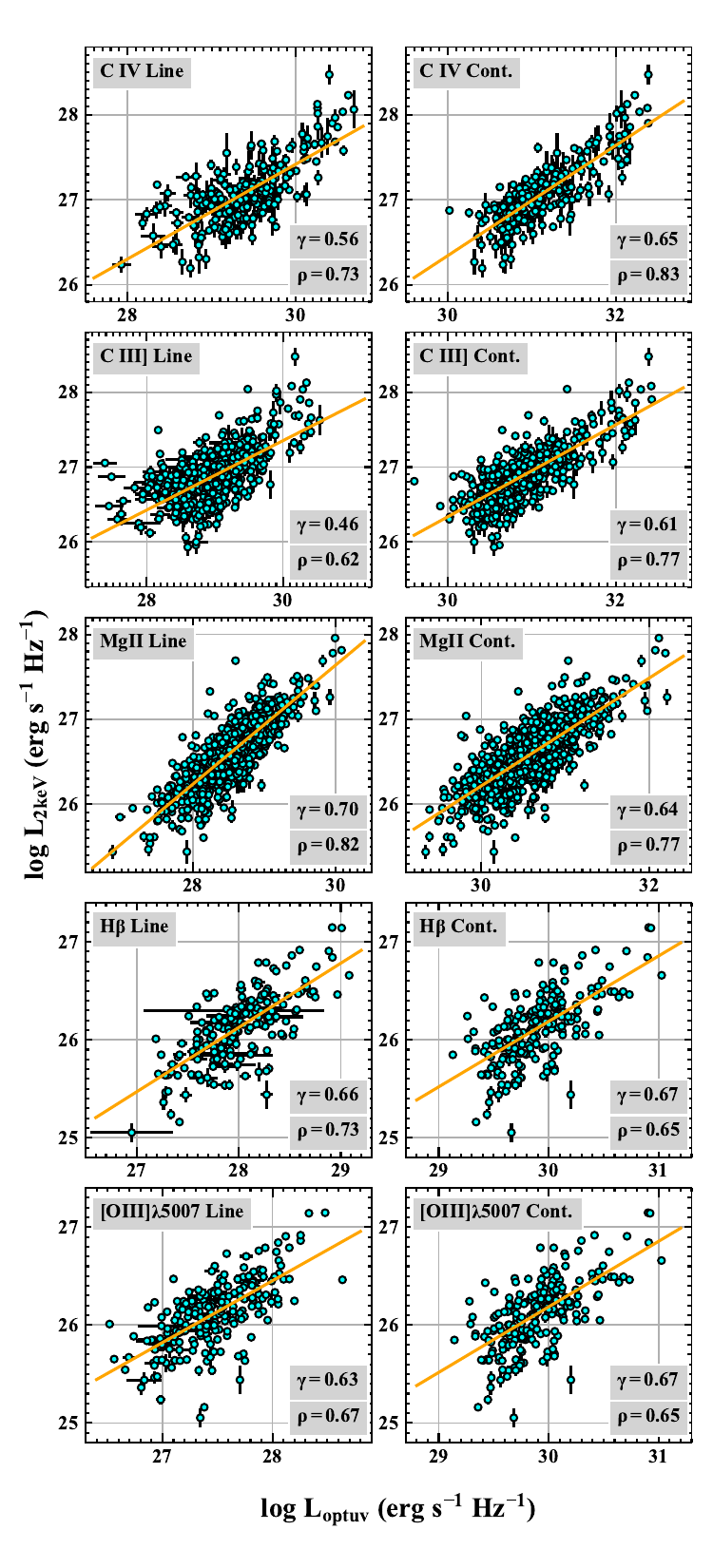} &
\includegraphics[trim=0.0in 0.2in 0.0in 0.1in, clip=1, scale=0.65]{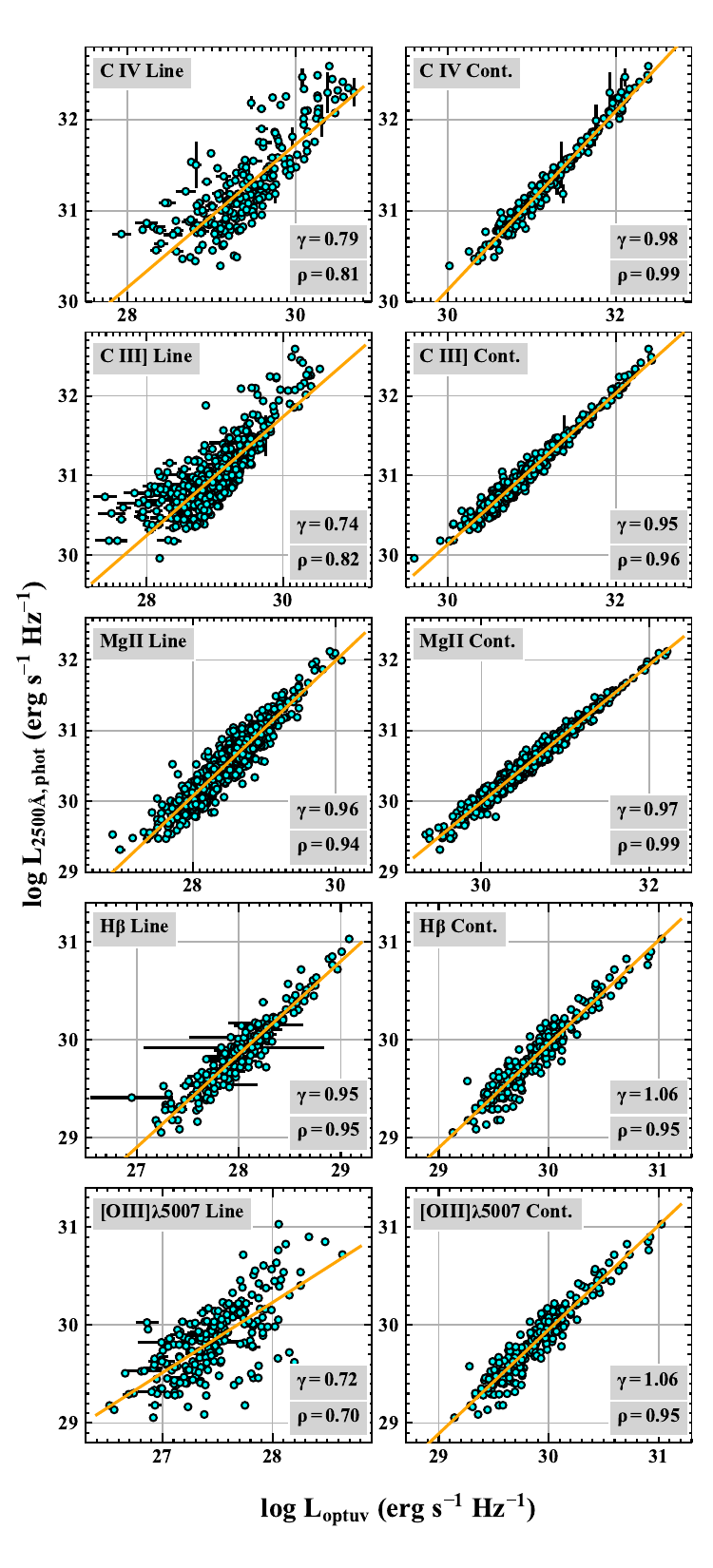} \\
\end{tabular}
\caption{Correlations in the left two columns are between the X-ray luminosity at 2 keV and a series of optical/UV emission lines and their underlying continua. The orange line indicates the regression line found by LINMIX. $\gamma$ and $\rho$ labeled in every panel are the correlation's slope and the Pearson's correlation coefficient. The right two columns show the correlations with the UV photometric luminosity at 2500 \AA.}
\label{fig-linecor}
\end{figure*}

\begin{figure*}
\centering
\begin{tabular}{cc}
\includegraphics[trim=0.2in 0.0in 0.0in 0.1in, clip=1, scale=0.65]{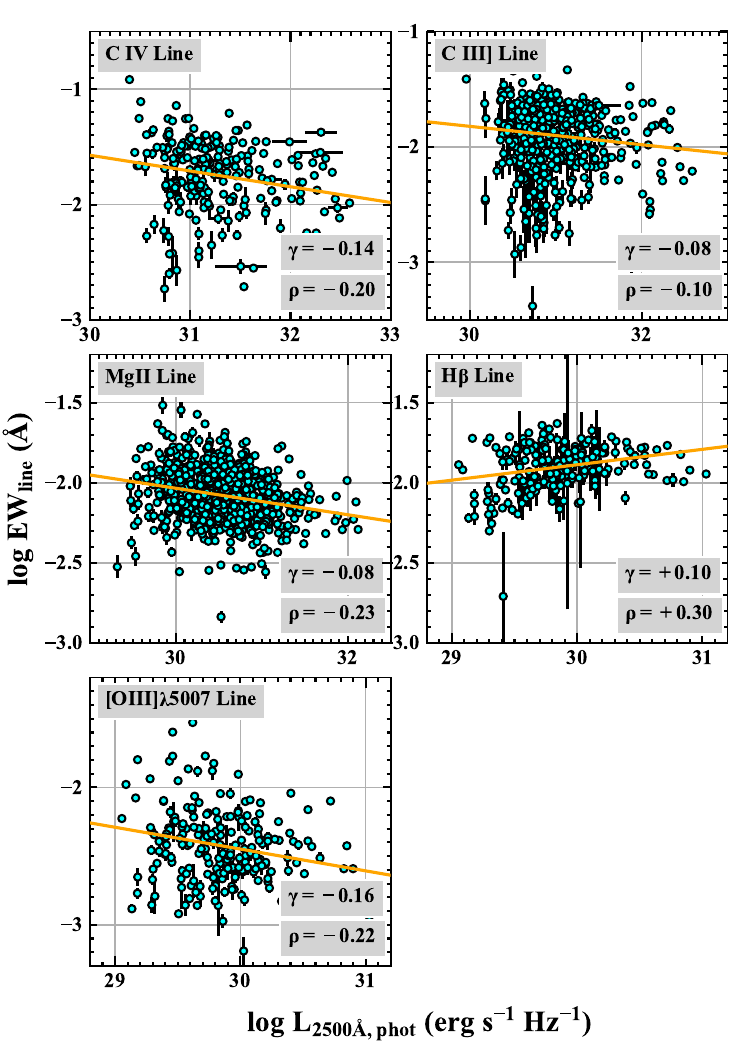} \\
\end{tabular}
\caption{Baldwin effect of different emission lines. The orange line indicates the regression line found by LINMIX. $\gamma$ and $\rho$ labeled in every panel are the correlation's slope and the Pearson's correlation coefficient. }
\label{fig-baldwin}
\end{figure*}



\bsp	
\label{lastpage}
\end{document}